\documentclass[sigconf,nonacm]{acmart}
\usepackage{float}
\usepackage{placeins}
\usepackage{booktabs}
\usepackage{array}
\usepackage{dcolumn}
\newcolumntype{F}{D{/}{/}{-1}}
\newcolumntype{G}{D{.}{.}{3.1}}
\newcolumntype{H}{D{.}{.}{4.1}}
\usepackage{accsupp}
\newcommand{\cnumhex}[1]{\ifcase#1\or 2460\or 2461\or 2462\or 2463\or 2464\fi}
\DeclareRobustCommand{\cnum}[1]{\BeginAccSupp{method=hex,unicode,ActualText=\cnumhex{#1}}\textcircled{#1}\EndAccSupp{}}
\usepackage{graphicx}
\usepackage{algorithm}
\usepackage{algpseudocode}
\usepackage{amsmath}
\usepackage{tikz}
\usetikzlibrary{arrows.meta,positioning,calc,decorations.pathreplacing}
\definecolor{esgaccent}{HTML}{1F5F9E}
\newtheorem{theorem}{Theorem}

\newtheorem{lemma}{Lemma}

\newcommand{\cand}{\mathcal{C}}
\newcommand{\gtrue}{G}
\newcommand{\ghat}{\hat{G}}
\newcommand{\gcal}{\tilde{G}}
\newcommand{\what}{\hat{W}}
\newcommand{\wcal}{\tilde{W}}
\newcommand{\slack}{s}

\newcommand{\ahat}{\hat{a}}
\newcommand{\viol}{\mathrm{viol}}
\newcommand{\METHOD}{\textsc{ESGuard}}
\newcommand{\POINT}{\textsc{Point}}
\newcommand{\MARGIN}{\textsc{Margin}}
\newcommand{\ORACLE}{\textsc{Candidate Oracle}}

\newcommand{\allow}{\beta}
\setcopyright{none}
\begin{document}
\title{Bias-Aware Guardrail Selection Under Network Interference}
\author{Minkyoung Kim}
\affiliation{\department{Graduate School of Information}\institution{Yonsei University}\city{Seoul}\country{Republic of Korea}}
\author{Hyunjung Byun}
\affiliation{\department{Graduate School of Information}\institution{Yonsei University}\city{Seoul}\country{Republic of Korea}}
\author{Beakcheol Jang}
\authornote{Corresponding author.}
\affiliation{\department{Graduate School of Information}\institution{Yonsei University}\city{Seoul}\country{Republic of Korea}}
\renewcommand{\shortauthors}{Kim, Byun, and Jang}
\begin{abstract}
Web platforms use randomized experiment logs to choose budget-feasible treatment allocations while limiting losses to a designated group of units. Under network interference, each allocation requires expected outcomes at treatment-exposure pairs that may be sparsely observed. Existing safe policy selection methods require valid inference for each candidate, but pooling nearby exposures to improve estimability can introduce allocation-specific bias that variance-only bounds omit. We propose exposure-stratified guardrail selection (\METHOD{}), which aggregates unit-level bias bounds into a bias bound on each candidate's guardrail effect. It incorporates these bounds into simultaneous lower confidence bounds to select an allocation or retain the status quo. Under randomized rounds, a correct exposure mapping, and valid curvature and slope bounds, the conditional bias bound holds in finite samples. With independent rounds and additional regularity conditions, the unconditional probability of selecting an unsafe allocation is asymptotically at most the target level. We evaluate \METHOD{} on a simulator built from New York City ride-hailing trips. With six months of simulated rounds, we compare estimators that pool nearby exposures on the same logs and candidate sets. Adding the bias bound raises simultaneous coverage from below the nominal 95 percent level to at least 95 percent for every estimator in this comparison. In the underserved-only scenario, \METHOD{} violates the guardrail in 0.06 percent of its selections, compared with 16 percent for the point-estimate rule under the same gain ranking, and abstains in about a quarter of replications. Omitted spillovers can invalidate the protection. Code is available at \url{https://anonymous.4open.science/r/esguard-4D2E/}.
\end{abstract}
\maketitle
\section{Introduction}
\label{sec:intro}

A ride-hailing platform may offer driver incentives to increase service in selected zones. With a limited budget, it must decide which zones receive them. We call the chosen set of zones an allocation. Incentives attract drivers to treated zones but can draw them from neighboring zones \cite{luo2024policy,shi2023multiagent}, so the allocation that maximizes total trips can reduce service where riders wait longest. The platform sets a guardrail limiting this loss to a stated tolerance \cite{schultzberg2026risk,wu2026marketplace}, weighing platform economics against unequal access in the gig economy. Similar decisions arise when marketplaces promote connected sellers or social platforms allocate features across regions.

We study how a platform uses its randomized experiment logs to choose a budget-feasible allocation from a finite candidate set, or retain the status quo, while controlling the probability of an unsafe selection. The log consists of repeated rounds, time slots as in switchback experiments \cite{bojinov2023switchback}. In each round, treatment is randomized independently across the units of a fixed graph, a design studied in Web research on platform interference \cite{brennan2025symbiosis,gui2015network,li2022interference}. Under an exposure mapping, a unit's expected outcome depends on its own treatment and its exposure, here the treated share of its neighbors \cite{aronow2017interference}. Randomized rounds identify each unit's expected outcome at the treatment-exposure pairs they record. The platform, however, must evaluate the allocation it plans to deploy. This requires each guarded unit's expected outcome at the treatment-exposure pair that allocation induces, and the experiment rarely records some of these pairs. For example, with six months of simulated rounds, exact-exposure estimates exist for only 17 to 19 percent of candidates.

Existing safe policy selection methods, including high-confidence policy improvement, Learn then Test, and CSPI-MT, control selection error through tests or confidence bounds that require valid inference for each candidate \cite{thomas2015hcpi,angelopoulos2025learn,cho2025cspimt}. In network logs, however, exact-exposure estimates at the pairs a candidate induces may be unavailable \cite{aronow2017interference,forastiere2021interference}, while pooling nearby exposures introduces candidate-dependent bias. Simultaneous inference accounts for sampling uncertainty across candidates, but a variance-only band around pooled estimates does not account for this bias. Bias-aware inference provides tools to bound smoothing error \cite{armstrong2018optimal,armstrong2020simple}. Applying them here, however, requires tracking the weights and exposure gaps of each unit-level fit and aggregating target-baseline errors into a bound for each allocation's guardrail.

\begin{figure}[t]
\centering
\begin{tikzpicture}[x=1cm, y=1cm, font=\footnotesize, >={Latex[length=1.2mm,width=1.0mm]},
  main/.style={align=center, inner sep=1pt},
  sub/.style={font=\scriptsize, text=black, align=center, inner sep=1pt},
  flow/.style={->, draw=black, line width=0.5pt},
  zone/.style={circle, draw=black, line width=0.55pt, minimum size=4.1mm, inner sep=0pt},
  trt/.style={zone, fill=esgaccent!70},
  prot/.style={zone, draw=black, line width=0.8pt, fill=black!8, minimum size=3.96mm}]
\def\CW{8.3}
\node[font=\scriptsize\bfseries, anchor=north west] at (0,3.35) {(a)};
\begin{scope}[shift={(1.3,2.3)}]
\node[prot] (c) at (0,0) {\scriptsize $i$};
\foreach \k/\t in {0/1,60/1,120/1,180/0,240/1,300/0}{
  \ifnum\t=1 \node[trt] (n\k) at (\k:0.68) {}; \else \node[zone] (n\k) at (\k:0.68) {}; \fi
  \draw[black!80, line width=0.55pt] (c) -- (n\k);}
\foreach \a/\b in {0/60,60/120,120/180,180/240,240/300,300/0}{\draw[black!30, line width=0.3pt] (n\a) -- (n\b);}
\node[sub, anchor=north] at (0,-0.97) {4 of 6 neighbors treated};
\end{scope}
\begin{scope}[shift={(2.95,1.42)}, y=0.85cm]
\def\X{4.9}
\fill[esgaccent!9] (0,0) rectangle (1.805,1.78);
\draw[black, line width=0.6pt] (0,0) -- (\X,0);
\draw[black, line width=0.4pt] (0,0) -- (0,1.78);
\node[sub, rotate=90, anchor=south] at (-0.02,0.89) {Outcome};
\foreach \v in {0.1667,0.3333,0.6667,0.8333}{\draw[black, line width=0.4pt] ({\X*\v},0) -- ({\X*\v},-0.05);}
\foreach \v/\l in {0/0,0.5/{1/2},1/1}{\draw[black, line width=0.6pt] ({\X*\v},0) -- ({\X*\v},-0.08) node[below, font=\scriptsize, inner sep=1pt] {$\l$};}
\node[sub, anchor=north] at ({\X*0.77},-0.07) {Exposure $e_i$};
\draw[black, dotted, line width=0.4pt] (3.267,0) -- (3.267,1.78);
\draw[black, line width=0.35pt, domain=0:1, samples=60, smooth] plot ({\X*\x},{0.45+0.95*(exp(-2*\x)-0.135335)/(1-0.135335)});
\draw[esgaccent, line width=0.45pt] (0,1.388) -- (1.805,0.794);
\draw[esgaccent, line width=0.45pt, dash pattern=on 1.6pt off 1.2pt] (1.805,0.794) -- (3.267,0.312);
\foreach \y in {1.22,1.34,1.46,1.58}{\filldraw[fill=black!35, draw=black!80, line width=0.35pt] (0,\y) circle (0.36mm);}
\foreach \x/\y in {0.817/0.97,0.817/1.09,0.817/1.21,1.633/0.75,1.633/0.87,1.633/0.98}{\filldraw[fill=black!35, draw=black!80, line width=0.35pt] (\x,\y) circle (0.38mm);}
\fill[black] (3.267,0.591) circle (0.39mm);
\node[draw=esgaccent, fill=white, line width=0.45pt, minimum size=0.88mm, inner sep=0pt] at (3.267,0.312) {};
\draw[black, line width=0.32pt, decorate, decoration={brace, amplitude=1.9pt}] (3.345,0.552) -- (3.345,0.298);
\node[sub, anchor=west, inner sep=0.5pt] at (3.43,0.335) {$|\mathrm{Bias}| \le b_i$};
\path (1.805,0.794) -- (3.267,0.312) node[pos=0.48, sloped, below, sub, text=esgaccent, inner sep=0.8pt] {Expected line fit};
\node[sub, anchor=south west] at (3.32,0.645) {True mean};
\node[sub] at (1.05,1.56) {Pooled bin};
\node[font=\scriptsize, anchor=west, inner sep=1pt] at (3.31,1.56) {$e_i(a) = 4/6$};
\node[sub, anchor=south] at ({\X*0.5},1.80) {Calibration rounds with $i$ untreated};
\end{scope}
\node[font=\scriptsize\bfseries, anchor=north west] at (0,0.62) {(b)};
\begin{scope}[shift={(0.45,-0.05)}]
\def\W{5.0}
\fill[esgaccent!10] (\W/2,0) rectangle (\W,0.42);
\draw[black, line width=0.5pt] (0,0) rectangle (\W,0.42) (\W/2,0) -- (\W/2,0.42);
\node[main] at (\W/4,0.21) {Training half};
\node[main] at (3*\W/4,0.21) {Calibration half};
\node[sub, anchor=south] at (\W/2,0.45) {Log of $R$ randomized rounds};
\node[main, anchor=north] (tr) at (\W/4,-0.28) {\cnum{1} Candidate set};
\node[sub, anchor=north] (tr2) at (tr.south) {from fitted gain and\\guardrail surfaces};
\node[main, anchor=north] (ca) at (3*\W/4,-0.28) {\cnum{2} Estimates and\\bias bounds\\{\scriptsize unit bounds summed into $\allow(a)$}\\\cnum{3} Simultaneous bounds};
\node[sub, anchor=north] (ca2) at (ca.south) {$\ell_\alpha(a) = \gcal(a) - \allow(a) - c_\alpha \hat{\nu}(a)$};
\draw[flow] (\W/4,0) -- (tr.north);
\draw[flow] (3*\W/4,0) -- (ca.north);
\draw[flow] (tr.east) -- (ca.west |- tr.east);
\node[main, draw=black, line width=0.5pt, inner sep=2.5pt, anchor=west] (sel) at (\W+0.42,0.21) {\cnum{4} Select $\max \wcal(a)$\\s.t.\ $\ell_\alpha(a) \ge -\slack$};
\node[main, text depth=0.25ex, anchor=north] (dep) at ([xshift=-0.7cm]sel.south |- 0,-0.5) {Deploy};
\node[main, text depth=0.25ex, anchor=north] (dec) at ([xshift=0.7cm]sel.south |- 0,-0.5) {Abstain};
\node[sub, anchor=north] at (dep.south) {if one passes\\with $\wcal > 0$};
\node[sub, anchor=north] at (dec.south) {otherwise};
\coordinate (turn) at (\W+0.2,0);
\draw[flow] (ca.east) -- (ca.east -| turn) |- (sel.west);
\draw[flow] ([xshift=-0.7cm]sel.south) -- (dep.north);
\draw[flow] ([xshift=0.7cm]sel.south) -- (dec.north);
\end{scope}
\end{tikzpicture}
\caption{(a) Candidate $a$ (filled zones) leaves guarded zone $i$ (bold outline) untreated with four of six neighbors treated, exposure $e_i(a) = 4/6$. The line fitted to the pooled bin (shaded) is extrapolated to $e_i(a)$, where its expected value (square) misses the unobserved true mean (dot). Under a curvature bound, $b_i$ bounds this gap. (b) The training half generates candidates. On the calibration half, each candidate's lower bound $\ell_\alpha(a)$ subtracts the bias bound $\allow(a)$ and the simultaneous band $c_\alpha \hat{\nu}(a)$ from its guardrail estimate $\gcal(a)$. A candidate passes if $\ell_\alpha(a) \ge -\slack$ for tolerance $\slack$ and is ranked by its gain estimate $\wcal(a)$.}
\Description{Panel a shows a guarded unit with four of six neighbors treated and its calibration outcomes against exposure, with the true mean curve, the line fitted to the pooled exposure bin and extrapolated to the target exposure, and the bias between them at that exposure. Panel b shows the training and calibration split, candidate generation, exposure-binned estimation, simultaneous bounds, and the decision to deploy or abstain.}
\label{fig:pipeline}
\end{figure}
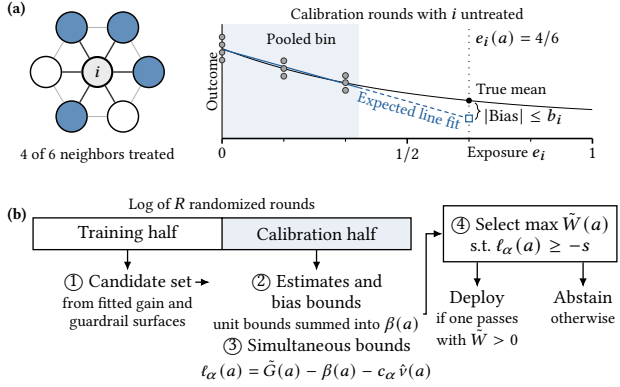

We propose exposure-stratified guardrail selection (\METHOD{}, Figure~\ref{fig:pipeline}). A training half generates candidate allocations, including partial allocations that leave budget unspent. On held-out calibration rounds, \METHOD{} merges rare exposures into neighboring bins and fits local lines. Conditional on the logged design, each fitted value is a weighted sum of outcomes, allowing a closed-form bias bound under response smoothness bounds \cite{armstrong2018optimal}. \METHOD{} aggregates these bounds over each candidate's guardrail contrasts and subtracts them when constructing simultaneous lower confidence bounds. It selects the passing candidate with the highest positive held-out gain, or retains the status quo.

We evaluate \METHOD{} on a simulator built from 3.1 million New York City ride-hailing trips, with known candidate guardrails. Spillover losses affect either all zones or only underserved zones. In the underserved-only scenario, the highest-gain allocation can be unsafe. In the main comparison, every smoothing or coarsening estimator tested falls below the nominal 95 percent coverage without the bias bound and reaches it with the bound, with different gain-abstention trade-offs. With six months of simulated rounds, \METHOD{} violates in 0.06 percent of its underserved-only selections (2 of 3,571), versus 16 percent for the point-estimate rule under the same gain ranking, and abstains in about a quarter of replications. It has no violations in the all-zone scenario.

Our contributions are:
\begin{itemize}
\item \textbf{Bias-aware allocation selection.} We develop \METHOD{} to select budgeted allocations under a group-level guardrail. It incorporates candidate-specific bounds on the bias from merging and extrapolation over sparse exposures into simultaneous inference, and retains the status quo when no candidate passes with positive estimated gain.
\item \textbf{Bias and selection guarantees.} Under randomized rounds, a correct exposure mapping, and valid curvature and slope bounds fixed before calibration, we bound each candidate's conditional bias for every number of rounds and binning outcome. With the additional conditions of Theorem~\ref{thm:band}, the unconditional probability of selecting an unsafe allocation is at most $\alpha + o(1)$ as the number of rounds grows, whichever passing candidate is selected.
\item \textbf{Controlled evaluation.} On the same logs and candidate sets, we compare exposure estimators under a common band procedure. We then compare safe-selection rules, including a CSPI-MT screening adaptation, under a common exposure estimator. We also compare against point and margin rules and examine performance under five omitted spillover structures and on a contagion testbed.
\end{itemize}

\section{Related Work}
\label{sec:related}

\paragraph{Safe policy selection}
High-confidence policy improvement \cite{thomas2015hcpi}, built on high-confidence off-policy evaluation \cite{thomas2015hcope}, and the Seldonian template \cite{thomas2019seldonian} test a candidate on held-out data and propose nothing if it fails. Learn then Test \cite{angelopoulos2025learn} controls the probability that any unsafe configuration passes within a finite set. CSPI-MT \cite{cho2025cspimt} evaluates threshold policies against a status quo with simultaneous bounds, while SNPL \cite{cho2025snpl} controls multiple guardrail outcomes. Conformal policy control \cite{prinster2026cpc} shows that selecting an aggressive candidate that appears safe favors estimates that understate its risk, a winner's curse \cite{lee2018winners,andrews2024inference}. Related approaches study abstention \cite{sawarni2025abstention}, chance constraints \cite{jia2025bayesian}, and constrained ranking and selection \cite{andradottir2010fully}. For advertising marketplaces, \citet{shekhar2026support} bound reserve-price policies under a replay estimand that excludes interference, and \citet{shekhar2026launch} require interference-aware validation before launch. A single change deployed only if all guardrails pass non-inferiority tests needs no multiplicity adjustment of its significance level \cite{schultzberg2026risk}. \METHOD{} supplies guardrail estimates and bias bounds for allocations whose effects depend on shared network exposures.

\paragraph{Estimation under interference}
Prior work studies targeting and allocation under interference with regret guarantees and capacity, strategic, fairness, or platform-scale constraints \cite{viviano2025targeting,kitagawa2023sequential,caljon2026interference,sahoo2025competing,kim2026fair,vandenakker2025profit,karlsson2025qini}. Other work studies budget-constrained targeting at platform scale without interference \cite{ai2022lbcf}. Exposure mappings \cite{aronow2017interference,forastiere2021interference} define the estimand used here, with extensions for differentiated neighbor effects \cite{lin2026gite} and transport across networks \cite{du2026transport}. Under misspecification, exposure-based estimators target an expected exposure effect \cite{savje2024causal}, and partial identification gives bounds on direct and spillover effects \cite{schroder2026misspecified}. Under approximate neighborhood interference, effects of distant treatments decay with distance \cite{leung2022causal}. \citet{loomba2026offpolicy} transport a Bernoulli design to a counterfactual policy under assumed interference.

Exact-exposure Horvitz--Thompson and H\'ajek estimates use rounds at the exposure of interest, which may be absent from a finite log. Smoothing exposure as a dose connects this problem to dose-response estimation \cite{hirano2004propensity,kennedy2017nonparametric}. The within-bin line is a degree-one partitioning regression, as in binned scatterplots \cite{cattaneo2024binscatter}. Bias-aware inference adjusts intervals for the largest smoothing bias allowed by a curvature bound \cite{armstrong2018optimal,armstrong2020simple}, including with discrete regressors \cite{kolesar2018inference}. \METHOD{} applies this reasoning per unit and own treatment, aggregates the bounds at each allocation's induced exposures, and carries them into simultaneous inference.

\paragraph{Platform experiment design}
Guardrails are standard in online controlled experiments \cite{kohavi2020trustworthy}. Designs that address interference include switchbacks \cite{bojinov2023switchback}, graph cluster randomization \cite{ugander2013graph}, and clustered switchbacks \cite{jia2023clustered}. Cluster randomization reduced interference bias in an Airbnb pricing meta-experiment \cite{holtz2025reducing}. Related work studies launch tests \cite{johari2024interference}, decision analysis for switchback experiments \cite{ni2025decision}, design selection \cite{shekhar2026designs}, and ramps that detect interference \cite{han2023detecting} or bound cumulative harm \cite{li2023phased}.

\section{Problem Formulation}
\label{sec:setting}

\paragraph{Units, allocations, and exposure}
Units $i = 1, \dots, n$ are nodes of an observed graph with adjacency matrix $A$. In the running example, units are taxi zones and edges link neighboring zones. An allocation $a \in \{0,1\}^n$ treats units with $a_i = 1$, subject to $\sum_i a_i \le B$. The budget need not be spent in full. Unit $i$'s exposure is $e_i(a) = \sum_j A_{ij} a_j / \sum_j A_{ij}$, the treated share of its neighbors. We assume its expected outcome depends on the allocation only through its own treatment and exposure, $(a_i, e_i(a))$. Two allocations that give unit $i$ the same own treatment and the same treated share of neighbors therefore give it the same expected outcome, whichever neighbors are treated. This is the exposure-mapping assumption \cite{aronow2017interference,forastiere2021interference}. \citet{savje2024causal} studies misspecified exposure mappings. We call this the one-hop mapping because it includes direct neighbors only.

\paragraph{Gain and guardrail}
Each unit records an outcome $y_i$, here served trips per slot. The guarded group $P$ contains the units the guardrail covers, here underserved zones where riders wait longest. Let $\mu_i(t, e)$ denote unit $i$'s expected outcome under own treatment $t$ and exposure $e$. The exposure-mapping assumption makes this quantity well defined. In the randomized log below, it equals $\mathbb{E}[y_i \mid a_i = t, e_i = e]$. Unit $i$'s expected change relative to no treatment is $\Delta_i(a) = \mu_i(a_i, e_i(a)) - \mu_i(0, 0)$. Summing these changes over all units gives the allocation's gain, and summing over $P$ gives its guardrail effect:
\begin{equation}
W(a) = \sum_{i=1}^{n} \Delta_i(a), \qquad \gtrue(a) = \sum_{i \in P} \Delta_i(a),
\label{eq:wg}
\end{equation}
The platform seeks to maximize $W(a)$ over budget-feasible allocations subject to the guardrail constraint $\gtrue(a) \ge -\slack$. The tolerance $\slack > 0$ is the accepted loss on $P$, a non-inferiority margin \cite{schultzberg2026risk}. An allocation with $\gtrue(a) < -\slack$ is unsafe.

In the example, gain measures trip volume, a platform experiment metric \cite{johari2022experimental,bojinov2023switchback}, rather than economic surplus. Every treated zone receives the same incentive, so the treatment budget also bounds total payment, a capacity constraint \cite{kitagawa2018who}. Subtracting payment would shift all allocations of one size equally.

\paragraph{Logged rounds}
Neither $W$ nor $\gtrue$ is observed. The platform has $R$ independent logged rounds. In each round $r$, every unit is independently treated with probability $\rho$, and all outcomes are recorded. A round thus contains one random assignment and its outcome vector. In the example, rounds are 30-minute slots with independent randomization across zones and slots. This differs from a market-wide switchback that treats every zone alike. We target within-slot effects. Steady-state effects of a persistent incentive, including supply responses over about two months \cite{hall2023reequilibrate}, are different estimands \cite{johari2022experimental,wager2021experimenting}.

\paragraph{Selection rules and the safety target}
A selection rule uses the log to choose an allocation $\ahat$ or abstain. Abstention retains the status quo $\emptyset$, with zero gain and guardrail change. The unconditional violation probability is
\begin{equation}
\viol = \Pr\bigl[\gtrue(\ahat) < -\slack,\ \ahat \text{ selected}\bigr],
\label{eq:viol}
\end{equation}
where probability is over the logged rounds. The rule meets the target at level $\alpha$ if $\viol \le \alpha$. Among logs that lead to a selection, the violation rate is $\Pr[\gtrue(\ahat) < -\slack \mid \ahat \text{ selected}]$. When selection has positive probability, this equals $\viol$ divided by that probability and can exceed $\alpha$ even when the unconditional target holds. Section~\ref{sec:method} bounds \eqref{eq:viol} asymptotically, and the experiments report both rates and abstention.

\section{Method}
\label{sec:method}

We randomly split the log into a training half and a calibration half of $R_{\mathrm{cal}} = R/2$ rounds. The training half generates the candidates and fixes the smoothness bounds. The calibration half estimates candidate guardrails and gains, constructs the guardrail bounds, and optionally checks residuals. The guarantee conditions on the training half and its candidate set, as in inference after sample splitting \cite{rinaldo2019bootstrapping}. Figure~\ref{fig:pipeline}b and Algorithm~\ref{alg:esguard} summarize the procedure.

\subsection{Candidate generation}
\label{sec:candidates}

We generate candidates that vary in both budget use and the trade-off between gain and the guardrail. Least-squares surfaces fitted on the training half estimate gain ($\what$) and guardrail effect ($\ghat$). Both are linear in unit covariates, own treatment, and exposure (Appendix~\ref{app:design}). Three allocation families use these surfaces. The Lagrangian path greedily adds the unit maximizing $\what(a) + \lambda \ghat(a)$ until the budget is spent, for a grid of $\lambda$ including $0$. The guardrail path uses the same greedy search subject to $\ghat(a) \ge -\slack/m$ for several $m \ge 1$, but this training-half constraint carries no guarantee. Partial allocations follow the $\lambda = 0$ path but stop at budget fractions $f \in \{1/8, 1/4, 1/2, 3/4\}$. They let smaller allocations pass when full-budget ones are unsafe.

The guarantee applies to any finite candidate set generated without calibration data. Gain is judged relative to that set, as policy-class comparisons are in empirical welfare maximization \cite{kitagawa2018who}. Forgone gain has two sources: the generator's gap between the best feasible allocation and the best safe candidate, and the selection rule's gap between that candidate and its chosen output.

\subsection{Exposure-binned estimates and bias bounds}
\label{sec:instrument}

For each unit in $P$, a candidate requires an expected outcome at its induced own treatment and exposure, which we call an evaluation point. Under the exposure mapping, calibration rounds with that own treatment inform the point through their recorded exposures, regardless of the remaining assignment. We therefore group rounds by own treatment and exposure \cite{aronow2017interference}, without matching the full neighborhood assignment, whose probability decreases exponentially with neighborhood size. Randomization identifies the conditional mean at each treatment and exposure, so estimation within these groups does not require inverse assignment-probability weights.

For each unit $i$ and own treatment $t \in \{0,1\}$, we build bins using the recorded design only. \cnum{1} Start with one bin per recorded exposure among rounds with $a^r_i=t$. \cnum{2} While at least two bins remain and one has fewer than $n_{\min}=15$ rounds, merge the bin with the fewest rounds with its less populated neighbor in exposure order. Ties select the lower exposure or lower neighbor. \cnum{3} Each bin spans its smallest to largest recorded exposure. Assign evaluation points between bins to the lower bin, and points outside the recorded range to the nearest end bin.

To estimate $\mu_i(t,e)$, including the baseline $\mu_i(0,0)$, fit outcome on exposure within the assigned bin and evaluate the line at $e$. A bin with fewer than three rounds or only one exposure uses its mean. If no round has own treatment $t$, use the mean of all of the unit's calibration outcomes and an infinite bias bound, so a candidate requiring that point cannot pass. The held-out guardrail estimate is
\begin{equation}
\gcal(a) = \sum_{i \in P} \bigl( \hat{\mu}_i(a_i, e_i(a)) - \hat{\mu}_i(0, 0) \bigr).
\label{eq:strat}
\end{equation}

Merging gains sampling precision at the cost of approximation bias, and points beyond recorded support require extrapolation. Relative to a bin mean, the local line removes the first-order bias from the exposure gap but leaves a second-order remainder. Because binning uses only treatments and exposures, conditioning on this design fixes both the bins and the fitting weights. Each fitted value is then $\hat{\mu}_i(t,e)=\sum_{r\in S}w_r(e)y^r_i$ (Appendix~\ref{app:biasaware}). Under $|\partial_e^2\mu_i(t,\cdot)|\le M_i$, its absolute bias is bounded by \citep[eq.~(11)]{armstrong2018optimal}
\begin{equation}
b_i(t, e) = \frac{M_i}{2} \sum_{r \in S} |w_r(e)|\, (e^r_i - e)^2 .
\label{eq:biasbound}
\end{equation}
For a fitted line, $b_i(t,e)$ equals the worst-case bias over the Taylor class of \citet{armstrong2018optimal}, which contains the curvature class by Taylor's theorem. It is zero for a single-value bin evaluated at that value and accounts for exposure gaps in merged or extrapolated bins. A mean bin adds $L_i|\bar e_S-e|$, where $L_i$ bounds $|\partial_e\mu_i(t,\cdot)|$ (Appendix~\ref{app:biasaware}). Summing target and baseline bounds over the guarded group gives the candidate's bias bound
\begin{equation}
\allow(a)=\sum_{i\in P:\,(a_i,e_i(a))\neq(0,0)}\bigl[b_i(a_i,e_i(a))+b_i(0,0)\bigr],
\label{eq:allow}
\end{equation}
which omits units whose evaluation point is the baseline, where the contrast in \eqref{eq:strat} is exactly zero. It bounds the bias of $\gcal(a)$ for every binning outcome.

Smoothness bounds must be fixed before calibration. Choosing them from the same outcomes used for inference need not preserve coverage \cite{armstrong2020simple}. We set rule-of-thumb values for $M_i$ and $L_i$ from global quadratic fits in exposure on the training half, separately for each unit and own treatment, with the own treatment suppressed in the notation \cite{armstrong2020simple,imbens2019optimized}. For a fitted quadratic $b_0 + b_1 e + b_2 e^2$, $M_i = 2|b_2|$ is its absolute second derivative and $L_i = \max(|b_1|, |b_1 + 2 b_2|)$ its maximum absolute slope over $[0,1]$ (Appendix~\ref{app:biasaware}). These values are fixed before calibration but are not guaranteed to bound the true derivatives. Section~\ref{sec:biasbound} distinguishes their empirical performance from the theorem's conditions and reports sensitivity at two and four times their values \cite{imbens2019optimized}.

\subsection{Simultaneous lower bounds}
\label{sec:band}

Selecting among estimates can invalidate a separate lower bound for each candidate: the selected estimate tends to have an optimistic error. Simultaneous bounds address this by holding for all candidates together at a chosen confidence level, such as 95 percent \cite{montielolea2019simultaneous,cho2025cspimt}. On that event, the selected candidate's bound holds.

Candidate estimates use shared outcomes and are correlated. The band preserves this dependence. Each held-out estimate is a weighted sum of the calibration outcomes, $\gcal(a) = \sum_r \sum_i c^r_i(a)\, y^r_i$. Let $u^r(a) = \sum_i c^r_i(a)\, \hat{\varepsilon}^r_i$ be the score of round $r$, with $\hat{\varepsilon}^r_i$ the leverage-scaled residual of its bin fit, and $\hat{\nu}(a) = (\sum_r u^r(a)^2)^{1/2}$ the sandwich standard error over rounds. With $\xi^r_b$ independent standard normal, $b = 1, \dots, B_{\mathrm{boot}}$, and the bins held fixed,
\begin{equation}
t_b = \max_{a \in \cand} \frac{\sum_r \xi^r_b\, u^r(a)}{\hat{\nu}(a)}, \qquad c_\alpha = \hat{Q}_{1-\alpha}\bigl(t_1, \dots, t_{B_{\mathrm{boot}}}\bigr),
\label{eq:band}
\end{equation}
with $\hat{Q}_{1-\alpha}$ the empirical $(1-\alpha)$-quantile. The lower bound is
\begin{equation}
\ell_\alpha(a) = \gcal(a) - \allow(a) - c_\alpha\, \hat{\nu}(a) \qquad \text{for all } a \in \cand \text{ at once}.
\label{eq:bound}
\end{equation}
Scores are summed within rounds to retain dependence among units. The maximum over candidates gives simultaneous inference: a one-sided version of the sup-$t$ band \cite{montielolea2019simultaneous}, as in CSPI-MT \cite{cho2025cspimt} and resampling-based max-$T$ testing \cite{westfall1993resampling,romano2005stepwise,white2000reality}. Any test that bounds the probability that some unsafe candidate passes can replace the band, such as the fixed-sequence test of Learn then Test \cite{angelopoulos2025learn} or a Bonferroni band (Appendix~\ref{app:nyc}).

\subsection{Selection and guarantees}
\label{sec:select}

\begin{algorithm}[t]
\caption{\METHOD{}}
\label{alg:esguard}
\small
\begin{algorithmic}[1]
\Require log $\{(a^r, y^r)\}_{r=1}^{R}$, budget $B$, tolerance $\slack$, level $\alpha$, $n_{\min}$, $B_{\mathrm{boot}}$
\Ensure an allocation to deploy, or abstention
\State Split the rounds at random into a training half and a calibration half
\State \cnum{1} On the training half, fit $\what$, $\ghat$, $M_i$, $L_i$ and generate $\cand$
\State \cnum{2} On the calibration half, build the exposure bins, then compute $\gcal(a)$ \eqref{eq:strat}, $\wcal(a)$, and $\allow(a)$ \eqref{eq:allow} for every $a \in \cand$
\State \cnum{3} Compute $u^r(a)$, $\hat{\nu}(a)$, and $c_\alpha$ \eqref{eq:band}, then $\ell_\alpha(a) \gets \gcal(a) - \allow(a) - c_\alpha \hat{\nu}(a)$
\State \cnum{4} $\mathcal{A} \gets \{a \in \cand : \ell_\alpha(a) \ge -\slack\}$
\State \Return the $a \in \mathcal{A}$ with the largest $\wcal(a)$ if that gain is positive and the optional residual check does not reject, and abstain otherwise
\end{algorithmic}
\end{algorithm}

The rule selects
\begin{equation}
\ahat = \arg\max_{a \in \cand \cup \{\emptyset\}} \wcal(a) \quad \text{subject to} \quad \ell_\alpha(a) \ge -\slack,
\label{eq:select}
\end{equation}
where $\wcal(a)$ uses the estimator in \eqref{eq:strat} summed over all units. The status quo has $\wcal(\emptyset)=0$, always passes, and wins ties. An allocation is therefore deployed only if its bound meets the tolerance and its estimated gain is positive. Otherwise, the rule abstains and retains the status quo, as in \citet{cho2025cspimt}. The guarantee protects the guardrail, not the gain.

If $\ell_\alpha(a) \le \gtrue(a)$ for every candidate, any passing candidate satisfies $\gtrue(a) \ge \ell_\alpha(a) \ge -\slack$ and is safe. An unsafe selection thus requires an unsafe candidate to pass, which in turn requires a simultaneous coverage failure (Lemma~\ref{lem:events}, Appendix~\ref{app:events}). This event inclusion permits ranking candidates by calibration-half gain: the guarantee holds whichever passing candidate is selected.

\begin{theorem}[Bias and selection guarantees]
\label{thm:band}
Hold the training half, $\cand$, and the bounds $M_i$, $L_i$ fixed, and condition on the own treatments and exposures recorded in the calibration half. Assume:
\begin{enumerate}
\item[(A1)] the rounds are randomized and the exposure mapping of Section~\ref{sec:setting} holds;
\item[(A2)] $|\partial_e^2 \mu_i(t, \cdot)| \le M_i$ for every $i \in P$ and $t$, and $|\partial_e \mu_i(t, \cdot)| \le L_i$ where a bin contributes its mean;
\item[(A3)] rounds are independent given the design, each with an outcome covariance over $P$ whose eigenvalues are bounded and bounded away from zero;
\item[(A4)] each round's contribution to every estimate has bounded kurtosis, and no round carries a non-vanishing share of any candidate's variance.
\end{enumerate}
Then under (A1) and (A2), (i) $|\mathbb{E}[\gcal(a) \mid \text{design}] - \gtrue(a)| \le \allow(a)$ for every $a \in \cand$ and every binning outcome, and under (A1) to (A4), (ii) the rule \eqref{eq:select} satisfies $\viol \le \alpha + o(1)$ on the fixed graph as the calibration half and $B_{\mathrm{boot}}$ grow, whichever candidate with $\ell_\alpha(a) \ge -\slack$ it selects.
\end{theorem}

\noindent Appendix~\ref{app:biasaware} gives the proof and an extension to other linear smoothers, estimators whose fitted values are weighted sums of outcomes with weights fixed by the design and the training half. The bias argument uses weights fixed by the design that reproduce constants. Weights that also reproduce lines remove the first-order term. The kernel local-linear fit reproduces lines, whereas coarsened H\'ajek reproduces constants only and requires a term in $L_i$. The variance conditions are also needed. For the kernel fit, part (ii) additionally requires vanishing residual smoothing error.

Part (i) is a finite-sample bias bound, whereas part (ii) is an asymptotic safety guarantee. Part (ii) is conditional on the training half and the calibration design, and its probability does not condition on whether a selection occurs. When each evaluation point matches its single-value bin, $\allow(a)=0$ and the lower bound reduces to the band without a bias bound. On a fixed graph, increasing calibration counts eventually produces such bins.

\paragraph{Residual check for omitted spillovers}
Before selection, an optional diagnostic checks whether calibration residuals are associated with two spillover variables that the one-hop mapping omits: the two-hop treated share and the previous round's exposure. It regresses the within-bin residuals on these covariates and permutes the residuals within unit, own treatment, and bin. The check-gated rule abstains if either permutation $p$-value is below $\alpha/2$. This follows randomization tests for omitted spillovers \cite{athey2018exact,basse2019randomization}, and Appendix~\ref{app:nyc} reports its empirical size. It can detect only spillovers associated with these covariates, and non-rejection does not establish that the mapping is correct.

\section{Experimental Setup}
\label{sec:experiments}

\paragraph{Data and network}
We use public New York City Taxi and Limousine Commission High Volume For-Hire Vehicle records \cite{nyctlc} from October 2024: weekdays, 16:00 to 20:00, grouped into 30-minute slots (about 3.1 million trips). For each zone, we compute mean trips per slot $d_i$, their slot-to-slot standard deviation $\nu_i$, and median request-to-pickup wait $w_i$. We remove zones with fewer than one trip per slot and the airport zone outside the city. The graph is the largest component of the union of 5-nearest-neighbor links between zone centroids. It contains 232 zones and 691 edges, with degrees 5 to 10. The guarded group $P$ comprises the 46 zones with the longest median waits.

\paragraph{Outcome model}
In round $r$ under assignment $a$, zone $i$ serves
\begin{equation}
y_i = d_i + \theta d_i a_i - \gamma \theta d_i \kappa_i h\bigl(e_i(a)\bigr) + \nu_i \epsilon_i, \qquad \epsilon_i \sim N(0, 1),
\label{eq:dgp}
\end{equation}
with $\theta = 0.10$, $\kappa_i = \mathrm{clip}(w_i / \mathrm{median}(w), 0.5, 2)$, and $h(e)=1-\exp(-3e)$, so $h(0)=0$ and $h(1)\approx0.95$. The incentive raises local trips by 10 percent. Treated neighbors impose a loss scaled by interference strength $\gamma$ and scarcity $\kappa_i$, with larger losses where waits are longer. These terms are motivated by observed supply responses to fare increases \cite{hall2023reequilibrate} and repositioning toward higher-paying areas \cite{lu2018surge}. Because public data do not identify the lift, we halve and double it in the sensitivity analysis (Section~\ref{sec:biasbound}).

The tolerance is $\slack=\delta_s\sum_{i\in P}d_i$, a fraction $\delta_s$ of the underserved zones' 3,279 trips per slot. Demand, noise scales, and waits ($d_i$, $\nu_i$, $w_i$) come from the records, whereas treatment effects and spillovers ($\theta$, $\gamma$, $\delta_s$, $h$) are chosen. Equation~\eqref{eq:dgp} gives $W$ and $\gtrue$ in closed form, allowing exact violation counts. Its main specification uses the same one-hop exposure as the estimator. Section~\ref{sec:sensitivity} and Appendix~\ref{app:nyc} examine spillovers outside that mapping and other response shapes. This combines real structure with simulated outcomes \cite{ma2022hypergraph,chen2024doubly}, as in ride-hailing simulators built on New York City trip records \cite{farias2022markovian,shirani2025validate}.

\paragraph{Spillover scenarios}
In the all-zone scenario, spillover losses apply to every zone. Unsafe allocations then tend to reduce total gain as well, so gain ranking alone often avoids them. In the underserved-only scenario, the loss in \eqref{eq:dgp} is zero for $i\notin P$, so the highest-gain allocation can be unsafe: the conflict the guardrail addresses.

Each scenario crosses $\gamma\in\{0.5,1,1.5,2\}$, $\delta_s\in\{0.003,0.01,0.03\}$, and treatment rates $\rho\in\{0.1,0.2\}$ with budgets $B=\mathrm{round}(\rho n)\in\{23,46\}$. $R\in\{160,480,960\}$ rounds represent about one, three, and six months of weekday-evening slots. A setting is one parameter combination, and a replication is one independently simulated log at that setting. We run 20 replications per setting at $R=160$ and 200 at $R\in\{480,960\}$, totaling 10,080 per scenario. Corresponding replications share assignments and noise across scenarios, while settings have independent seeds. At $R=160$, $\rho=0.1$, fewer than 15 treated calibration rounds per zone are expected, so this setting is a boundary diagnostic. Longer experiments, $R\in\{1920,3840\}$ (about one and two years), use 20 replications per setting in both scenarios. Unless stated, results use $R=960$.

Candidate sets contain 8 to 16 allocations. The default training surface has one shared exposure coefficient (Section~\ref{sec:candidates}). A second generator fits a separate exposure coefficient for each unit.

\paragraph{Evaluation design}
All methods receive the same logs and initial candidate sets, and only \ORACLE{} uses the true effects. Appendix~\ref{app:nyc} specifies their selection rules and exceptions. Default bias bounds use the rule-of-thumb values of $M_i$ and $L_i$ (Section~\ref{sec:instrument}). Rows under true derivative bounds in Table~\ref{tab:allow} instead use the outcome model's bounds.

\emph{(1) How does the exposure estimator affect performance?} We hold the band procedure fixed and replace the within-bin line with a bin mean, H\'ajek-exact, Horvitz--Thompson, coarsened H\'ajek, or a kernel local-linear fit. Every row ranks candidates by the within-bin line's gain estimate, so rows differ only in the guardrail estimate. H\'ajek-exact averages outcomes over rounds with exactly the evaluation exposure, and Horvitz--Thompson weights those rounds by inverse design probability. Both leave candidates without a usable bound when a required exposure is unrecorded. Coarsened H\'ajek uses the mean at the nearest lower recorded exposure. Following \citet{kennedy2017nonparametric}, the kernel fit is a local-linear smoother per unit and own treatment with leave-one-out bandwidth selection on the training half. It uses an Epanechnikov kernel.

\emph{(2) Does the bias bound restore simultaneous coverage?} We compare the band without a bias bound with bounds based on one, two, and four times the rule-of-thumb values.

\emph{(3) How do selection rules trade off safety, gain, and abstention?} We compare \METHOD{} with \POINT{} (pass if the estimate meets the tolerance), \MARGIN-$m$ (apply $\slack/m$ to the training surface), fixed-sequence testing (candidates tested in a training-half order until the first failure), Bonferroni, a Seldonian procedure, and a CSPI-MT screening adaptation \cite{cho2025cspimt} that chooses on the training half the subset of candidates to test. \ORACLE{} returns the best truly safe candidate or the status quo.

\paragraph{Metrics and statistical tests}
Before the runs, we fixed three questions: whether \POINT{} and \MARGIN-$m$ exceed a 5 percent target at $R=960$, whether \METHOD{} meets it over pooled $R=480$ and $R=960$ runs, and whether its abstention falls from $R=160$ to $R=960$ at each $\rho$. Each target assessment combines the selections across $\gamma$, $\delta_s$, $\rho$, and replications and uses the share that violate the guardrail. This violation rate among selections is at least as large as the unconditional violation rate. There are 4,800 replications per row at $R\in\{480,960\}$ and 480 elsewhere.

We test exceedance with a one-sided binomial test of $H_0\colon p\le0.05$ at the 5 percent level. We assess whether the target is met using a one-sided 95 percent Clopper--Pearson upper bound (CP bound) below 0.05. Violation probabilities vary across settings, but at the thresholds used here, \citet[Theorem~4]{hoeffding1956distribution} keeps both tests at the 5 percent level for their average (Appendix~\ref{app:nyc}). We also report unconditional violations per setting, the scope of the guarantee \cite{morris2019using}. Fisher's exact test assesses the fall in abstention.

All methods use $\alpha = 0.05$. Violations are reported as violating selections over selections, and abstention as the share of replications that abstain. Coverage is the share of replications in which every evaluated candidate's lower bound lies at or below its true guardrail effect, with nominal level 95 percent. Exact-exposure estimators evaluate only the candidates they define, and CSPI-MT's screen only those it tests. Gain is the selected allocation's true gain, averaged over all replications with zero for abstention. Violations over all replications estimate \eqref{eq:viol}.

\paragraph{Ethical use of data and code}
The trip records are public and contain no rider or driver identifiers, and the testbed keeps zone-level aggregates. Every treatment effect is simulated, and no experiment involved riders or drivers. Code, configurations, and the summary files behind every table and figure are at \url{https://anonymous.4open.science/r/esguard-4D2E/}.
\section{Results}
\label{sec:results}

\begin{table*}[t]
\caption{Main results at $R = 960$, pooled over settings. Upper block: estimators under the same band and gain ranking, without ($k = 0$) and with ($k = 1$) the bias bound. Lower block: decision rules. Violations: unsafe selections over selections, with a one-sided 95 percent Clopper--Pearson upper bound (CP bound). Coverage of the exact-exposure estimators and of the screen is over the candidates they evaluate. $^\dagger$Network adaptation (Appendix~\ref{app:nyc}).}
\label{tab:main}
\centering
\footnotesize
\setlength{\tabcolsep}{3pt}
\begin{tabular}{lcFcccHcFcccG}
\toprule
& & \multicolumn{5}{c}{All-zone} & & \multicolumn{5}{c}{Underserved-only} \\
\cmidrule(lr){3-7} \cmidrule(lr){9-13}
Method & $k$ & \multicolumn{1}{c}{Violations} & CP bound & Coverage & Abstention & \multicolumn{1}{c}{Gain} & & \multicolumn{1}{c}{Violations} & CP bound & Coverage & Abstention & \multicolumn{1}{c}{Gain} \\
\midrule
\multicolumn{13}{l}{\emph{Guardrail estimators under the same band}} \\
\METHOD{} (line) & 0 & 2/2803 & 0.002 & 0.938 & 0.416 & 59.5 & & 26/3697 & 0.010 & 0.937 & 0.230 & 214.4 \\
 & 1 & 0/2716 & 0.001 & 0.982 & 0.434 & 52.7 & & 2/3571 & 0.002 & 0.983 & 0.256 & 175.5 \\
\METHOD{} (mean) & 0 & 11/2898 & 0.006 & 0.690 & 0.396 & 68.1 & & 271/4057 & 0.074 & 0.691 & 0.155 & 279.2 \\
 & 1 & 0/2638 & 0.001 & 0.981 & 0.450 & 39.5 & & 0/3564 & 0.001 & 0.982 & 0.258 & 131.5 \\
H\'ajek-coarsened & 0 & 1/2805 & 0.002 & 0.919 & 0.416 & 60.4 & & 25/3708 & 0.009 & 0.918 & 0.228 & 221.1 \\
 & 1 & 0/2723 & 0.001 & 0.976 & 0.433 & 54.6 & & 2/3573 & 0.002 & 0.978 & 0.256 & 185.0 \\
Kernel local-linear & 0 & 1/2995 & 0.002 & 0.891 & 0.376 & 62.7 & & 31/4130 & 0.010 & 0.901 & 0.140 & 242.1 \\
 & 1 & 0/2260 & 0.001 & 0.999 & 0.529 & 35.3 & & 0/2929 & 0.001 & 0.999 & 0.390 & 115.1 \\
H\'ajek-exact & -- & 0/2912 & 0.001 & 0.949 & 0.393 & 37.7 & & 4/3962 & 0.002 & 0.951 & 0.175 & 141.2 \\
Horvitz--Thompson & -- & 1/2342 & 0.002 & 0.446 & 0.512 & 31.8 & & 330/3540 & 0.102 & 0.456 & 0.263 & 135.6 \\
\midrule
\multicolumn{13}{l}{\emph{Selection rules}} \\
\POINT{}, $\wcal$ & -- & 25/3216 & 0.011 & & 0.330 & 62.0 & & 755/4722 & 0.169 & & 0.016 & 307.2 \\
\MARGIN{}, $m = 2$ & -- & 2353/4800 & 0.502 & & 0.000 & -245.3 & & 2415/4800 & 0.515 & & 0.000 & 386.2 \\
Fixed sequence & 1 & 0/2572 & 0.001 & & 0.464 & 46.4 & & 4/3648 & 0.003 & & 0.240 & 167.4 \\
Bonferroni band & 1 & 0/2536 & 0.001 & 0.993 & 0.472 & 48.8 & & 0/3275 & 0.001 & 0.991 & 0.318 & 158.2 \\
Seldonian, $\eta = 1$ & -- & 0/2543 & 0.001 & & 0.470 & 53.5 & & 0/2840 & 0.001 & & 0.408 & 168.8 \\
CSPI-MT screen$^\dagger$ & 0 & 1/2809 & 0.002 & 0.948 & 0.415 & 59.2 & & 2/3895 & 0.002 & 0.948 & 0.189 & 203.5 \\
 & 1 & 0/2589 & 0.001 & 0.964 & 0.461 & 48.2 & & 1/3578 & 0.001 & 0.967 & 0.255 & 165.7 \\
\ORACLE{} & -- & 0/3126 & 0.001 & & 0.349 & 81.9 & & 0/4762 & 0.001 & & 0.008 & 324.8 \\
\bottomrule
\end{tabular}
\end{table*}

\subsection{Estimator comparison}
\label{sec:estimators}

Without the bias bound, every smoothing or coarsening estimator in the upper block of Table~\ref{tab:main} has coverage below the nominal 95 percent level. Coverage is 0.89 to 0.94 across the within-bin line, coarsened H\'ajek, and the kernel fit. The bin mean retains the first-order bias removed by the line: its coverage is 0.69, and 7 percent of its underserved-only selections violate.

Exact-exposure estimators avoid smoothing bias but cannot evaluate most candidates. H\'ajek-exact covers at about the nominal level on the 17 percent of candidates it defines, with gain 20 to 28 percent below \METHOD{} with the bias bound. Horvitz--Thompson defines 19 percent but covers in fewer than half of replications on that set. Conditional on recorded exposures, its expectation equals the mean scaled by the ratio of recorded to expected rounds (Appendix~\ref{app:nyc}). The residual band does not account for this conditional error. This result concerns the estimator--band pairing used here, rather than Horvitz--Thompson's validity over the randomized design.

In Table~\ref{tab:main}, adding the bias bound brings every smoothing and coarsening estimator tested to at least the nominal 95 percent level, with different gain and abstention costs. The bin mean covers in 0.98 but yields 25 percent less gain than the line in the underserved-only scenario. Coarsened H\'ajek has the same abstention as the line and slightly lower coverage: 0.976 and 0.978 versus 0.982 and 0.983. Its gain is higher by 9.5 trips per slot in the underserved-only scenario (paired standard error 0.9) and 1.9 in the all-zone scenario (0.4).

We use the within-bin line by default because it estimates every evaluation point from rounds with the same own treatment and permits extrapolation in both directions. The kernel fit leaves 3 percent of candidates undefined.

\subsection{Bias-bound sensitivity}
\label{sec:biasbound}

\begin{table}[t]
\caption{\METHOD{} by number of rounds $R$, underserved-only (all-zone in Appendix~\ref{app:nyc}). $k$ scales the rule-of-thumb $M_i$ and $L_i$, and $k = 0$ omits the bias bound. Unsafe passes (some unsafe candidate passes) and unsafe selections count replications out of 4,800 (480 at $R \ge 1920$).}
\label{tab:allow}
\centering
\footnotesize
\setlength{\tabcolsep}{3pt}
\begin{tabular}{rrcrrcr}
\toprule
& & & \multicolumn{2}{c}{Unsafe} & & \\
\cmidrule(lr){4-5}
$R$ & $k$ & Coverage & Passes & Selections & Abstention & Gain \\
\midrule
\multicolumn{7}{l}{\emph{Rule-of-thumb values}} \\
480 & 0 & 0.955 & 15 & 13 & 0.356 & 181.5 \\
 & 1 & 0.988 & 2 & 2 & 0.514 & 87.1 \\
960 & 0 & 0.937 & 37 & 26 & 0.230 & 214.4 \\
 & 1 & 0.983 & 3 & 2 & 0.256 & 175.5 \\
 & 2 & 0.985 & 1 & 0 & 0.265 & 156.9 \\
 & 4 & 0.986 & 1 & 0 & 0.272 & 135.7 \\
1920 & 0 & 0.921 & 3 & 1 & 0.115 & 246.8 \\
 & 1 & 0.969 & 0 & 0 & 0.119 & 224.7 \\
3840 & 0 & 0.929 & 3 & 1 & 0.040 & 268.2 \\
 & 1 & 0.963 & 1 & 0 & 0.040 & 259.0 \\
\midrule
\multicolumn{7}{l}{\emph{True derivative bounds}} \\
480 & -- & 0.983 & 2 & 2 & 0.449 & 121.1 \\
960 & -- & 0.981 & 2 & 2 & 0.249 & 187.3 \\
1920 & -- & 0.977 & 0 & 0 & 0.117 & 227.0 \\
3840 & -- & 0.971 & 1 & 0 & 0.040 & 258.2 \\
\bottomrule
\end{tabular}
\end{table}

In Table~\ref{tab:allow}, the bias bound keeps coverage at or above the nominal 95 percent level at every $R$, with a gain cost that decreases as $R$ grows. Up to six months, 44 percent of evaluation points lie in multi-exposure bins and 9 percent beyond recorded support. Without the bias bound, coverage is 0.955 at $R=480$ and 0.92 to 0.94 at larger $R$, and 26 selections violate at $R=960$. Merging bias persists over these values of $R$ while the sampling band narrows. With the bound, at most two selections violate at any $R$.

The cost is largest at small $R$, where bins are wide. At $R=480$, the bias bound raises abstention from 36 to 51 percent and halves gain. With one and two years of slots, abstention is unchanged and gain falls by 9 and 3 percent as bins narrow. The choice of $k$ follows a separate rule fixed before the runs. Pooling the three- and six-month runs within each scenario, the rule chooses $k=1$ if it raises abstention over $k=0$ by at most 0.10 in both scenarios, and $k=0$ otherwise. The rule sets one default for all runs and uses neither violations nor coverage. It selects $k=1$, with abstention increases of 0.07 and 0.09.

The rule-of-thumb values need not bound the true derivatives, as Theorem~\ref{thm:band} requires. At six months, $M_i$ bounds the largest curvature of $\mu_i(t,\cdot)$ for 48 percent of units in $P$, and $2M_i$ and $4M_i$ do so for 71 and 86 percent. Results using these values therefore describe empirical performance rather than establish the theorem's premise. The true derivative bounds in Table~\ref{tab:allow} instead set $M_i$ and $L_i$ to the outcome model's derivative suprema, which a platform would not know. No candidate's absolute bias $|\mathbb{E}[\gcal(a)\mid\text{design}]-\gtrue(a)|$ exceeds its bound in these runs, as part (i) predicts. Remaining coverage failures arise from the sampling band, and coverage is 0.97 to 0.98.

At $R=480$, true derivative bounds yield 121 trips per slot, compared with 87 for the rule-of-thumb values at $k=1$. At this $R$, the rule-of-thumb bias bound is nearly twice the bound from the true derivatives, in the median ratio over candidates. Conversely, at $R=960$ the absolute bias exceeds the rule-of-thumb bound for 0.3 to 0.4 percent of candidates. Most such estimates understate the guardrail, keeping the lower bound below the truth. Empirical coverage therefore remains 0.96 to 0.99. This favorable bias direction is not guaranteed. Under the convex response $h(e) = e^2$, the line's bias shifts toward the unsafe side. At $R=960$, adding the rule-of-thumb bias bound raises coverage from 0.84 to 0.98 and reduces replications with an unsafe pass from 16 of 600 to none. Under this convex response, abstention is 0.09 versus 0.08 and realized gain 107 versus 101 trips per slot, with versus without the bound (Appendix~\ref{app:nyc}).

Coverage and unsafe selection are distinct: Lemma~\ref{lem:events} bounds the latter by coverage failure. At $k=0$ and six months, coverage is 0.937, yet only 26 of 4,800 replications (0.5 percent) select an unsafe allocation.

\paragraph{Robustness to design choices}
With $n_{\min} = 30$ instead of 15, \METHOD{} violates in 1 of 10,521 selections over $R$ from 480 to 3840 and both scenarios, with coverage 0.98 to 0.99. With surfaces fitted per unit, it violates in 3 of 17,214 selections, with coverage 0.94 to 0.98. We also halve and double the lift $\theta$, rewire 5 to 20 percent of the edges of the graph the method uses while outcomes follow the true graph, use recorded residuals as noise, and use training fractions of 0.3 and 0.7. Coverage stays at 0.94 to 0.99 except at 20 percent rewired, where it falls to 0.88. That variant has the only violations among these variants, 2 in 206 selections. Appendix~\ref{app:nyc} reports further sensitivity results.

\begin{figure}[t]
\centering
\includegraphics[width=\columnwidth]{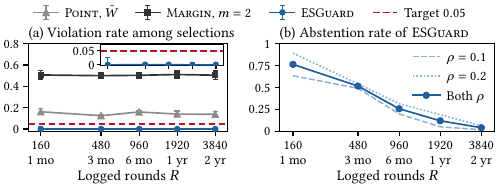}
\caption{Underserved-only scenario across $R$: (a) violation rate among selections with one-sided 95 percent Clopper--Pearson upper bounds (inset: \METHOD{}), (b) \METHOD{} abstention. Every $m$ gives the same \MARGIN{} curve.}
\Description{Panel a compares the violation rates among selections of the point, margin, and ESGuard rules across $R$, with confidence bounds and an inset for ESGuard. Panel b shows ESGuard abstention across $R$ and treatment rates in the underserved-only scenario.}
\label{fig:e2}
\end{figure}

\subsection{Selection rules and candidate generation}
\label{sec:platform}

In the underserved-only scenario, point and fixed-margin rules violate in at least 16 percent of selections, while all safety tests meet the target. With training-surface gain ranking, \POINT{} violates in 22 and 23 percent of selections across the two scenarios. Held-out gain ranking avoids most unsafe allocations in the all-zone scenario, where low guardrail effects tend to accompany low gain. It does not do so in the underserved-only scenario, where unsafe allocations can carry the most trips. \MARGIN{} violates in about half of its selections at every $m$. In every replication some candidate meets the training-surface margin, so no selection is forced.

Fixed-sequence testing abstains about as often as \METHOD{}, whereas Bonferroni and the Seldonian procedure abstain more. With the bias bound, the CSPI-MT screening adaptation also meets the target. It tests 2.5 and 3.6 candidates on average but yields 4.5 and 9.8 fewer trips per slot than \METHOD{} (paired standard errors 0.6 and 1.3). Without the bias bound, the adaptation reduces underserved-only violations to 2 of 3,895 selections, compared with 26 of 3,697 for the full band.

Candidate generation also determines the available gain. In the underserved-only scenario, more than half of replications contain an unsafe candidate with positive gain. \ORACLE{}, the best safe candidate or the status quo, bounds the gain obtainable by always selecting safely from this set. It achieves 59 percent of the gain of a greedy allocation built from true outcomes subject to the guardrail. \METHOD{} achieves 54 percent of \ORACLE{}'s gain in the underserved-only scenario and 64 percent in the all-zone scenario. The first gap reflects candidate generation, and the second reflects selection within the generated set. Underserved-only abstention falls from 76 percent with one month of slots to 4 percent with two years (Figure~\ref{fig:e2}b).

The trade-off changes with the generator. With per-unit fitted surfaces, 1 to 2 percent of candidates are unsafe at $R=960$, against 34 percent with the shared coefficient, and the point rule violates in at most 2.1 percent of selections in the underserved-only scenario. \METHOD{} selects in all but 0.3 percent of replications, with no violations, and retains 95 percent of the gain of \POINT{} under training ranking. The safety test's benefit and cost therefore depend on which candidates it must distinguish.

\subsection{Exposure-mapping misspecification}
\label{sec:sensitivity}

\begin{table}[t]
\caption{Spillover outside the one-hop mapping at $R = 960$ (1,200 replications per scenario, 480 under carryover), violations pooled over the two scenarios and the minimum coverage across them.}
\label{tab:sens}
\centering
\footnotesize
\setlength{\tabcolsep}{2pt}
\begin{tabular}{lFccF}
\toprule
& \multicolumn{2}{c}{\METHOD} & Check & \multicolumn{1}{c}{Check-gated} \\
\cmidrule(lr){2-3}
Spillover & \multicolumn{1}{c}{Violations} & Coverage & Rejection rate & \multicolumn{1}{c}{Violations} \\
\midrule
None (mapping holds) & 1/1576 & 0.986 & 0.05 & 1/1502 \\
Demand-weighted & 30/1673 & 0.705 & 0.05 & 29/1596 \\
Two-hop & 37/1888 & 0.803 & 0.55 & 3/977 \\
Distance decay & 161/2063 & 0.623 & 0.75 & 0/572 \\
Carryover, $c = 0.25$ & 10/761 & 0.819 & 0.64 & 0/323 \\
Carryover, $c = 0.5$ & 91/858 & 0.560 & 0.84 & 0/152 \\
\quad Lagged mapping & 2/287 & 0.948 & & \\
Trip flow & 1286/2300 & 0.127 & 0.36 & 601/1500 \\
\quad Flow mapping & 7/54 & 0.978 & & \\
\bottomrule
\end{tabular}
\end{table}

When the estimator keeps the one-hop mapping, omitted demand-weighted, two-hop, distance-decay, and carryover spillovers reduce \METHOD{}'s coverage to 0.56--0.82 (Table~\ref{tab:sens}). Under two-hop and carryover spillovers, unconditional violations increase with $R$ (Appendix~\ref{app:nyc}): specification bias remains while the band narrows. Adding previous-round treatment and exposure to the fit (the lagged mapping) restores coverage to 0.95 under carryover with $c=0.5$, with violations in 0.7 percent of selections.

Trip flows connect distant zones, creating mapping errors shared across many units. The limited dependence of specification errors on which exposure-estimator consistency relies \citep[Condition~4]{savje2024causal} may then fail. With the one-hop mapping, \METHOD{} violates in more than half of its selections, or 40 percent when gated by the residual check. Using trip-flow exposure (the flow mapping) restores nominal coverage, but the median band half-width is about 18 times the tolerance and abstention reaches 98 percent. Violations occur in 13 percent of selections (7 of 54), yet in only 0.3 percent of all replications. The guarantee targets this unconditional probability.

The residual check rejects in 5 percent of replications under the correct mapping and in 55 to 84 percent under two-hop, distance-decay, and carryover spillovers at $R=960$. It has little power against demand weighting, which neither diagnostic covariate tracks. Gating removes nearly all violations under those three spillovers but raises abstention to 59--84 percent. For trip flows it lowers the selected-only violation rate from 0.56 to 0.40, but for demand weighting it leaves the rate largely unchanged.

On the contagion testbed (Appendix~\ref{app:nyc}), where infection spreads beyond one hop and the one-hop mapping is an approximation, \METHOD{} violates in 2.4 percent of selections at 300 and 1,500 rounds (8 of 337), versus 20.5 percent for \POINT{} with training ranking (343 of 1,677). Its coverage with the bias bound falls from 0.997 at 300 rounds to 0.93 at 6,000.

\section{Discussion and Conclusion}
\label{sec:discussion}

\paragraph{Main findings}
Sparse exposure support creates a bias--precision trade-off in guardrail estimation. In the estimator comparison at $R=960$, adding candidate-specific bias bounds brings the within-bin line, coarsened H\'ajek, and kernel fit to at least nominal coverage. Gain and abstention trade-offs vary across estimators and selection rules.

\paragraph{Scope and extensions}
\METHOD{} answers one decision per log, which allocation to deploy in the next slots or whether to retain the status quo, and its guarantee applies to that decision, not jointly to a sequence of deployments. The same band construction can accommodate other weighted-sum estimators with valid bias bounds and the variance conditions of Theorem~\ref{thm:band}, such as those of \citet{loomba2026offpolicy}.

\paragraph{Limitations}
Treatment effects and spillovers are simulated on real demand, noise scales, graph structure, and underserved zones. The estimand captures within-slot effects, whereas persistent incentives can change market equilibrium \cite{wager2021experimenting}. The guarantee requires a correct exposure mapping, independent rounds, and valid smoothness bounds. Partial identification under a misspecified mapping \cite{schroder2026misspecified} could relax the first requirement. The rule-of-thumb values need not bound the true derivatives. Carryover from driver repositioning \cite{bojinov2023switchback,hu2022switchback,wu2026switchback} requires past assignments in the mapping. Surrogate-variable estimators \cite{chen2026efficient} may help represent such effects. Adaptively logged rounds lie outside the guarantee.

\bibliographystyle{ACM-Reference-Format}
\bibliography{refs}


\begin{thebibliography}{78}


\ifx \showCODEN    \undefined \def \showCODEN     #1{\unskip}     \fi
\ifx \showISBNx    \undefined \def \showISBNx     #1{\unskip}     \fi
\ifx \showISBNxiii \undefined \def \showISBNxiii  #1{\unskip}     \fi
\ifx \showISSN     \undefined \def \showISSN      #1{\unskip}     \fi
\ifx \showLCCN     \undefined \def \showLCCN      #1{\unskip}     \fi
\ifx \shownote     \undefined \def \shownote      #1{#1}          \fi
\ifx \showarticletitle \undefined \def \showarticletitle #1{#1}   \fi
\ifx \showURL      \undefined \def \showURL       {\relax}        \fi
\providecommand\bibfield[2]{#2}
\providecommand\bibinfo[2]{#2}
\providecommand\natexlab[1]{#1}
\providecommand\showeprint[2][]{arXiv:#2}

\bibitem[Ai et~al\mbox{.}(2022)]%
        {ai2022lbcf}
\bibfield{author}{\bibinfo{person}{Meng Ai}, \bibinfo{person}{Biao Li},
  \bibinfo{person}{Heyang Gong}, \bibinfo{person}{Qingwei Yu},
  \bibinfo{person}{Shengjie Xue}, \bibinfo{person}{Yuan Zhang},
  \bibinfo{person}{Yunzhou Zhang}, {and} \bibinfo{person}{Peng Jiang}.}
  \bibinfo{year}{2022}\natexlab{}.
\newblock \showarticletitle{{LBCF}: A Large-Scale Budget-Constrained Causal
  Forest Algorithm}. In \bibinfo{booktitle}{\emph{Proceedings of the ACM Web
  Conference 2022}}. \bibinfo{pages}{2310--2319}.
\newblock


\bibitem[Andrad{\'o}ttir and Kim(2010)]%
        {andradottir2010fully}
\bibfield{author}{\bibinfo{person}{Sigr{\'u}n Andrad{\'o}ttir} {and}
  \bibinfo{person}{Seong-Hee Kim}.} \bibinfo{year}{2010}\natexlab{}.
\newblock \showarticletitle{Fully Sequential Procedures for Comparing
  Constrained Systems via Simulation}.
\newblock \bibinfo{journal}{\emph{Naval Research Logistics}}
  \bibinfo{volume}{57}, \bibinfo{number}{5} (\bibinfo{year}{2010}),
  \bibinfo{pages}{403--421}.
\newblock


\bibitem[Andrews et~al\mbox{.}(2024)]%
        {andrews2024inference}
\bibfield{author}{\bibinfo{person}{Isaiah Andrews}, \bibinfo{person}{Toru
  Kitagawa}, {and} \bibinfo{person}{Adam McCloskey}.}
  \bibinfo{year}{2024}\natexlab{}.
\newblock \showarticletitle{Inference on Winners}.
\newblock \bibinfo{journal}{\emph{The Quarterly Journal of Economics}}
  \bibinfo{volume}{139}, \bibinfo{number}{1} (\bibinfo{year}{2024}),
  \bibinfo{pages}{305--358}.
\newblock


\bibitem[Angelopoulos et~al\mbox{.}(2025)]%
        {angelopoulos2025learn}
\bibfield{author}{\bibinfo{person}{Anastasios~N. Angelopoulos},
  \bibinfo{person}{Stephen Bates}, \bibinfo{person}{Emmanuel~J. Cand{\`e}s},
  \bibinfo{person}{Michael~I. Jordan}, {and} \bibinfo{person}{Lihua Lei}.}
  \bibinfo{year}{2025}\natexlab{}.
\newblock \showarticletitle{Learn Then Test: Calibrating Predictive Algorithms
  to Achieve Risk Control}.
\newblock \bibinfo{journal}{\emph{The Annals of Applied Statistics}}
  \bibinfo{volume}{19}, \bibinfo{number}{2} (\bibinfo{year}{2025}),
  \bibinfo{pages}{1641--1662}.
\newblock
\href{https://doi.org/10.1214/24-AOAS1998}{doi:\nolinkurl{10.1214/24-AOAS1998}}


\bibitem[Armstrong and Koles{\'a}r(2018)]%
        {armstrong2018optimal}
\bibfield{author}{\bibinfo{person}{Timothy~B. Armstrong} {and}
  \bibinfo{person}{Michal Koles{\'a}r}.} \bibinfo{year}{2018}\natexlab{}.
\newblock \showarticletitle{Optimal Inference in a Class of Regression Models}.
\newblock \bibinfo{journal}{\emph{Econometrica}} \bibinfo{volume}{86},
  \bibinfo{number}{2} (\bibinfo{year}{2018}), \bibinfo{pages}{655--683}.
\newblock


\bibitem[Armstrong and Koles{\'a}r(2020)]%
        {armstrong2020simple}
\bibfield{author}{\bibinfo{person}{Timothy~B. Armstrong} {and}
  \bibinfo{person}{Michal Koles{\'a}r}.} \bibinfo{year}{2020}\natexlab{}.
\newblock \showarticletitle{Simple and Honest Confidence Intervals in
  Nonparametric Regression}.
\newblock \bibinfo{journal}{\emph{Quantitative Economics}}
  \bibinfo{volume}{11}, \bibinfo{number}{1} (\bibinfo{year}{2020}),
  \bibinfo{pages}{1--39}.
\newblock


\bibitem[Aronow and Samii(2017)]%
        {aronow2017interference}
\bibfield{author}{\bibinfo{person}{Peter~M. Aronow} {and}
  \bibinfo{person}{Cyrus Samii}.} \bibinfo{year}{2017}\natexlab{}.
\newblock \showarticletitle{Estimating Average Causal Effects under General
  Interference, with Application to a Social Network Experiment}.
\newblock \bibinfo{journal}{\emph{Annals of Applied Statistics}}
  \bibinfo{volume}{11}, \bibinfo{number}{4} (\bibinfo{year}{2017}),
  \bibinfo{pages}{1912--1947}.
\newblock


\bibitem[Athey et~al\mbox{.}(2018)]%
        {athey2018exact}
\bibfield{author}{\bibinfo{person}{Susan Athey}, \bibinfo{person}{Dean Eckles},
  {and} \bibinfo{person}{Guido~W. Imbens}.} \bibinfo{year}{2018}\natexlab{}.
\newblock \showarticletitle{Exact $p$-Values for Network Interference}.
\newblock \bibinfo{journal}{\emph{J. Amer. Statist. Assoc.}}
  \bibinfo{volume}{113}, \bibinfo{number}{521} (\bibinfo{year}{2018}),
  \bibinfo{pages}{230--240}.
\newblock


\bibitem[Basse et~al\mbox{.}(2019)]%
        {basse2019randomization}
\bibfield{author}{\bibinfo{person}{Guillaume~W. Basse}, \bibinfo{person}{Avi
  Feller}, {and} \bibinfo{person}{Panos Toulis}.}
  \bibinfo{year}{2019}\natexlab{}.
\newblock \showarticletitle{Randomization Tests of Causal Effects Under
  Interference}.
\newblock \bibinfo{journal}{\emph{Biometrika}} \bibinfo{volume}{106},
  \bibinfo{number}{2} (\bibinfo{year}{2019}), \bibinfo{pages}{487--494}.
\newblock


\bibitem[Bojinov et~al\mbox{.}(2023)]%
        {bojinov2023switchback}
\bibfield{author}{\bibinfo{person}{Iavor Bojinov}, \bibinfo{person}{David
  Simchi-Levi}, {and} \bibinfo{person}{Jinglong Zhao}.}
  \bibinfo{year}{2023}\natexlab{}.
\newblock \showarticletitle{Design and Analysis of Switchback Experiments}.
\newblock \bibinfo{journal}{\emph{Management Science}} \bibinfo{volume}{69},
  \bibinfo{number}{7} (\bibinfo{year}{2023}), \bibinfo{pages}{3759--3777}.
\newblock


\bibitem[Brennan et~al\mbox{.}(2025)]%
        {brennan2025symbiosis}
\bibfield{author}{\bibinfo{person}{Jennifer Brennan}, \bibinfo{person}{Yahu
  Cong}, \bibinfo{person}{Yiwei Yu}, \bibinfo{person}{Lina Lin},
  \bibinfo{person}{Yajun Peng}, \bibinfo{person}{Changping Meng},
  \bibinfo{person}{Ningren Han}, \bibinfo{person}{Jean Pouget-Abadie}, {and}
  \bibinfo{person}{David~M. Holtz}.} \bibinfo{year}{2025}\natexlab{}.
\newblock \showarticletitle{Reducing Symbiosis Bias through Better {A/B} Tests
  of Recommendation Algorithms}. In \bibinfo{booktitle}{\emph{Proceedings of
  the ACM on Web Conference 2025}}. \bibinfo{pages}{3702--3715}.
\newblock


\bibitem[Caljon et~al\mbox{.}(2026)]%
        {caljon2026interference}
\bibfield{author}{\bibinfo{person}{Daan Caljon}, \bibinfo{person}{Jente
  Van~Belle}, \bibinfo{person}{Jeroen Berrevoets}, {and}
  \bibinfo{person}{Wouter Verbeke}.} \bibinfo{year}{2026}\natexlab{}.
\newblock \showarticletitle{Optimizing Treatment Allocation in the Presence of
  Interference}.
\newblock \bibinfo{journal}{\emph{European Journal of Operational Research}}
  \bibinfo{volume}{328}, \bibinfo{number}{2} (\bibinfo{year}{2026}),
  \bibinfo{pages}{620--632}.
\newblock


\bibitem[Cattaneo et~al\mbox{.}(2024)]%
        {cattaneo2024binscatter}
\bibfield{author}{\bibinfo{person}{Matias~D. Cattaneo},
  \bibinfo{person}{Richard~K. Crump}, \bibinfo{person}{Max~H. Farrell}, {and}
  \bibinfo{person}{Yingjie Feng}.} \bibinfo{year}{2024}\natexlab{}.
\newblock \showarticletitle{On Binscatter}.
\newblock \bibinfo{journal}{\emph{American Economic Review}}
  \bibinfo{volume}{114}, \bibinfo{number}{5} (\bibinfo{year}{2024}),
  \bibinfo{pages}{1488--1514}.
\newblock


\bibitem[Chen and Simchi-Levi(2026)]%
        {chen2026efficient}
\bibfield{author}{\bibinfo{person}{Hongyu Chen} {and} \bibinfo{person}{David
  Simchi-Levi}.} \bibinfo{year}{2026}\natexlab{}.
\newblock \showarticletitle{Efficient Switchback Experiments with Surrogate
  Variables: Estimation and Experimental Design}.
\newblock \bibinfo{journal}{\emph{Management Science}} \bibinfo{volume}{72},
  \bibinfo{number}{6} (\bibinfo{year}{2026}), \bibinfo{pages}{4854--4870}.
\newblock


\bibitem[Chen et~al\mbox{.}(2024)]%
        {chen2024doubly}
\bibfield{author}{\bibinfo{person}{Weilin Chen}, \bibinfo{person}{Ruichu Cai},
  \bibinfo{person}{Zeqin Yang}, \bibinfo{person}{Jie Qiao},
  \bibinfo{person}{Yuguang Yan}, \bibinfo{person}{Zijian Li}, {and}
  \bibinfo{person}{Zhifeng Hao}.} \bibinfo{year}{2024}\natexlab{}.
\newblock \showarticletitle{Doubly Robust Causal Effect Estimation under
  Networked Interference via Targeted Learning}. In
  \bibinfo{booktitle}{\emph{Proceedings of the 41st International Conference on
  Machine Learning}} \emph{(\bibinfo{series}{Proceedings of Machine Learning
  Research}, Vol.~\bibinfo{volume}{235})}. \bibinfo{pages}{6457--6485}.
\newblock


\bibitem[Chernozhukov et~al\mbox{.}(2013)]%
        {chernozhukov2013gaussian}
\bibfield{author}{\bibinfo{person}{Victor Chernozhukov}, \bibinfo{person}{Denis
  Chetverikov}, {and} \bibinfo{person}{Kengo Kato}.}
  \bibinfo{year}{2013}\natexlab{}.
\newblock \showarticletitle{Gaussian Approximations and Multiplier Bootstrap
  for Maxima of Sums of High-Dimensional Random Vectors}.
\newblock \bibinfo{journal}{\emph{The Annals of Statistics}}
  \bibinfo{volume}{41}, \bibinfo{number}{6} (\bibinfo{year}{2013}),
  \bibinfo{pages}{2786--2819}.
\newblock
\href{https://doi.org/10.1214/13-AOS1161}{doi:\nolinkurl{10.1214/13-AOS1161}}


\bibitem[Cho et~al\mbox{.}(2025a)]%
        {cho2025snpl}
\bibfield{author}{\bibinfo{person}{Brian~M. Cho}, \bibinfo{person}{Ana-Roxana
  Pop}, \bibinfo{person}{Ariel Evnine}, {and} \bibinfo{person}{Nathan Kallus}.}
  \bibinfo{year}{2025}\natexlab{a}.
\newblock \showarticletitle{{SNPL}: Simultaneous Policy Learning and Evaluation
  for Safe Multi-Objective Policy Improvement}.
\newblock \bibinfo{journal}{\emph{arXiv preprint arXiv:2503.12760}}
  (\bibinfo{year}{2025}).
\newblock


\bibitem[Cho et~al\mbox{.}(2025b)]%
        {cho2025cspimt}
\bibfield{author}{\bibinfo{person}{Brian~M. Cho}, \bibinfo{person}{Ana-Roxana
  Pop}, \bibinfo{person}{Kyra Gan}, \bibinfo{person}{Sam Corbett-Davies},
  \bibinfo{person}{Israel Nir}, \bibinfo{person}{Ariel Evnine}, {and}
  \bibinfo{person}{Nathan Kallus}.} \bibinfo{year}{2025}\natexlab{b}.
\newblock \showarticletitle{{CSPI-MT}: Calibrated Safe Policy Improvement with
  Multiple Testing for Threshold Policies}. In
  \bibinfo{booktitle}{\emph{Proceedings of the 31st ACM SIGKDD Conference on
  Knowledge Discovery and Data Mining}}. \bibinfo{pages}{165--176}.
\newblock


\bibitem[Du et~al\mbox{.}(2026)]%
        {du2026transport}
\bibfield{author}{\bibinfo{person}{Xiaojing Du}, \bibinfo{person}{Jiuyong Li},
  \bibinfo{person}{Lin Liu}, \bibinfo{person}{Debo Cheng},
  \bibinfo{person}{Jixue Liu}, {and} \bibinfo{person}{Thuc~Duy Le}.}
  \bibinfo{year}{2026}\natexlab{}.
\newblock \showarticletitle{Transportable Causal Effect Estimation across
  Networks under Interference}.
\newblock \bibinfo{journal}{\emph{arXiv preprint arXiv:2608.18932}}
  (\bibinfo{year}{2026}).
\newblock


\bibitem[Farias et~al\mbox{.}(2022)]%
        {farias2022markovian}
\bibfield{author}{\bibinfo{person}{Vivek~F. Farias}, \bibinfo{person}{Andrew~A.
  Li}, \bibinfo{person}{Tianyi Peng}, {and} \bibinfo{person}{Andrew Zheng}.}
  \bibinfo{year}{2022}\natexlab{}.
\newblock \showarticletitle{Markovian Interference in Experiments}. In
  \bibinfo{booktitle}{\emph{Advances in Neural Information Processing Systems
  35}}. \bibinfo{pages}{535--549}.
\newblock


\bibitem[Forastiere et~al\mbox{.}(2021)]%
        {forastiere2021interference}
\bibfield{author}{\bibinfo{person}{Laura Forastiere},
  \bibinfo{person}{Edoardo~M. Airoldi}, {and} \bibinfo{person}{Fabrizia
  Mealli}.} \bibinfo{year}{2021}\natexlab{}.
\newblock \showarticletitle{Identification and Estimation of Treatment and
  Interference Effects in Observational Studies on Networks}.
\newblock \bibinfo{journal}{\emph{J. Amer. Statist. Assoc.}}
  \bibinfo{volume}{116}, \bibinfo{number}{534} (\bibinfo{year}{2021}),
  \bibinfo{pages}{901--918}.
\newblock


\bibitem[Gui et~al\mbox{.}(2015)]%
        {gui2015network}
\bibfield{author}{\bibinfo{person}{Huan Gui}, \bibinfo{person}{Ya Xu},
  \bibinfo{person}{Anmol Bhasin}, {and} \bibinfo{person}{Jiawei Han}.}
  \bibinfo{year}{2015}\natexlab{}.
\newblock \showarticletitle{Network {A/B} Testing: From Sampling to
  Estimation}. In \bibinfo{booktitle}{\emph{Proceedings of the 24th
  International Conference on World Wide Web}}. \bibinfo{pages}{399--409}.
\newblock


\bibitem[Hall et~al\mbox{.}(2023)]%
        {hall2023reequilibrate}
\bibfield{author}{\bibinfo{person}{Jonathan~V. Hall}, \bibinfo{person}{John~J.
  Horton}, {and} \bibinfo{person}{Daniel~T. Knoepfle}.}
  \bibinfo{year}{2023}\natexlab{}.
\newblock \bibinfo{booktitle}{\emph{Ride-Sharing Markets Re-Equilibrate}}.
\newblock \bibinfo{type}{NBER Working Paper} 30883.
  \bibinfo{institution}{National Bureau of Economic Research}.
\newblock


\bibitem[Han et~al\mbox{.}(2023)]%
        {han2023detecting}
\bibfield{author}{\bibinfo{person}{Kevin Han}, \bibinfo{person}{Shuangning Li},
  \bibinfo{person}{Jialiang Mao}, {and} \bibinfo{person}{Han Wu}.}
  \bibinfo{year}{2023}\natexlab{}.
\newblock \showarticletitle{Detecting Interference in Online Controlled
  Experiments with Increasing Allocation}. In
  \bibinfo{booktitle}{\emph{Proceedings of the 29th ACM SIGKDD Conference on
  Knowledge Discovery and Data Mining}}. \bibinfo{pages}{661--672}.
\newblock


\bibitem[Hirano and Imbens(2004)]%
        {hirano2004propensity}
\bibfield{author}{\bibinfo{person}{Keisuke Hirano} {and}
  \bibinfo{person}{Guido~W. Imbens}.} \bibinfo{year}{2004}\natexlab{}.
\newblock \showarticletitle{The Propensity Score with Continuous Treatments}.
\newblock In \bibinfo{booktitle}{\emph{Applied Bayesian Modeling and Causal
  Inference from Incomplete-Data Perspectives}},
  \bibfield{editor}{\bibinfo{person}{Andrew Gelman} {and}
  \bibinfo{person}{Xiao-Li Meng}} (Eds.). \bibinfo{publisher}{Wiley},
  \bibinfo{pages}{73--84}.
\newblock


\bibitem[Hoeffding(1956)]%
        {hoeffding1956distribution}
\bibfield{author}{\bibinfo{person}{Wassily Hoeffding}.}
  \bibinfo{year}{1956}\natexlab{}.
\newblock \showarticletitle{On the Distribution of the Number of Successes in
  Independent Trials}.
\newblock \bibinfo{journal}{\emph{The Annals of Mathematical Statistics}}
  \bibinfo{volume}{27}, \bibinfo{number}{3} (\bibinfo{year}{1956}),
  \bibinfo{pages}{713--721}.
\newblock


\bibitem[Holtz et~al\mbox{.}(2025)]%
        {holtz2025reducing}
\bibfield{author}{\bibinfo{person}{David Holtz}, \bibinfo{person}{Felipe
  Lobel}, \bibinfo{person}{Ruben Lobel}, \bibinfo{person}{Inessa Liskovich},
  {and} \bibinfo{person}{Sinan Aral}.} \bibinfo{year}{2025}\natexlab{}.
\newblock \showarticletitle{Reducing Interference Bias in Online Marketplace
  Experiments Using Cluster Randomization: Evidence from a Pricing
  Meta-experiment on {Airbnb}}.
\newblock \bibinfo{journal}{\emph{Management Science}} \bibinfo{volume}{71},
  \bibinfo{number}{1} (\bibinfo{year}{2025}), \bibinfo{pages}{390--406}.
\newblock


\bibitem[Hu and Wager(2022)]%
        {hu2022switchback}
\bibfield{author}{\bibinfo{person}{Yuchen Hu} {and} \bibinfo{person}{Stefan
  Wager}.} \bibinfo{year}{2022}\natexlab{}.
\newblock \showarticletitle{Switchback Experiments under Geometric Mixing}.
\newblock \bibinfo{journal}{\emph{arXiv preprint arXiv:2209.00197}}
  (\bibinfo{year}{2022}).
\newblock


\bibitem[Imbens and Wager(2019)]%
        {imbens2019optimized}
\bibfield{author}{\bibinfo{person}{Guido Imbens} {and} \bibinfo{person}{Stefan
  Wager}.} \bibinfo{year}{2019}\natexlab{}.
\newblock \showarticletitle{Optimized Regression Discontinuity Designs}.
\newblock \bibinfo{journal}{\emph{The Review of Economics and Statistics}}
  \bibinfo{volume}{101}, \bibinfo{number}{2} (\bibinfo{year}{2019}),
  \bibinfo{pages}{264--278}.
\newblock


\bibitem[Jia et~al\mbox{.}(2023)]%
        {jia2023clustered}
\bibfield{author}{\bibinfo{person}{Su Jia}, \bibinfo{person}{Nathan Kallus},
  {and} \bibinfo{person}{Christina~Lee Yu}.} \bibinfo{year}{2023}\natexlab{}.
\newblock \showarticletitle{Clustered Switchback Designs for Experimentation
  Under Spatio-temporal Interference}.
\newblock \bibinfo{journal}{\emph{arXiv preprint arXiv:2312.15574}}
  (\bibinfo{year}{2023}).
\newblock


\bibitem[Jia et~al\mbox{.}(2026)]%
        {jia2025bayesian}
\bibfield{author}{\bibinfo{person}{Zeyang Jia}, \bibinfo{person}{Eli
  Ben-Michael}, {and} \bibinfo{person}{Kosuke Imai}.}
  \bibinfo{year}{2026}\natexlab{}.
\newblock \showarticletitle{Bayesian Safe Policy Learning with Chance
  Constrained Optimization: Application to Military Security Assessment during
  the {Vietnam} War}.
\newblock \bibinfo{journal}{\emph{Journal of the Royal Statistical Society
  Series A: Statistics in Society}} \bibinfo{volume}{189}, \bibinfo{number}{3}
  (\bibinfo{year}{2026}), \bibinfo{pages}{1448--1472}.
\newblock
\href{https://doi.org/10.1093/jrsssa/qnaf122}{doi:\nolinkurl{10.1093/jrsssa/qnaf122}}


\bibitem[Johari et~al\mbox{.}(2022)]%
        {johari2022experimental}
\bibfield{author}{\bibinfo{person}{Ramesh Johari}, \bibinfo{person}{Hannah Li},
  \bibinfo{person}{Inessa Liskovich}, {and} \bibinfo{person}{Gabriel~Y.
  Weintraub}.} \bibinfo{year}{2022}\natexlab{}.
\newblock \showarticletitle{Experimental Design in Two-Sided Platforms: An
  Analysis of Bias}.
\newblock \bibinfo{journal}{\emph{Management Science}} \bibinfo{volume}{68},
  \bibinfo{number}{10} (\bibinfo{year}{2022}), \bibinfo{pages}{7069--7089}.
\newblock


\bibitem[Johari et~al\mbox{.}(2024)]%
        {johari2024interference}
\bibfield{author}{\bibinfo{person}{Ramesh Johari}, \bibinfo{person}{Hannah Li},
  \bibinfo{person}{Anushka Murthy}, {and} \bibinfo{person}{Gabriel~Y.
  Weintraub}.} \bibinfo{year}{2024}\natexlab{}.
\newblock \showarticletitle{When Does Interference Matter? Decision-Making in
  Platform Experiments}.
\newblock \bibinfo{journal}{\emph{arXiv preprint arXiv:2410.06580}}
  (\bibinfo{year}{2024}).
\newblock


\bibitem[Karlsson et~al\mbox{.}(2025)]%
        {karlsson2025qini}
\bibfield{author}{\bibinfo{person}{Rickard Karlsson}, \bibinfo{person}{Bram
  van~den Akker}, \bibinfo{person}{Felipe Moraes}, \bibinfo{person}{Hugo~M.
  Proen{\c{c}}a}, {and} \bibinfo{person}{Jesse~H. Krijthe}.}
  \bibinfo{year}{2025}\natexlab{}.
\newblock \showarticletitle{Qini Curve Estimation under Clustered Network
  Interference}.
\newblock \bibinfo{journal}{\emph{arXiv preprint arXiv:2502.20097}}
  (\bibinfo{year}{2025}).
\newblock


\bibitem[Kennedy et~al\mbox{.}(2017)]%
        {kennedy2017nonparametric}
\bibfield{author}{\bibinfo{person}{Edward~H. Kennedy},
  \bibinfo{person}{Zongming Ma}, \bibinfo{person}{Matthew~D. McHugh}, {and}
  \bibinfo{person}{Dylan~S. Small}.} \bibinfo{year}{2017}\natexlab{}.
\newblock \showarticletitle{Non-parametric Methods for Doubly Robust Estimation
  of Continuous Treatment Effects}.
\newblock \bibinfo{journal}{\emph{Journal of the Royal Statistical Society:
  Series B (Statistical Methodology)}} \bibinfo{volume}{79},
  \bibinfo{number}{4} (\bibinfo{year}{2017}), \bibinfo{pages}{1229--1245}.
\newblock


\bibitem[Kim et~al\mbox{.}(2026)]%
        {kim2026fair}
\bibfield{author}{\bibinfo{person}{Raphael~C. Kim}, \bibinfo{person}{Rachel~C.
  Nethery}, \bibinfo{person}{Kevin~L. Chen}, {and} \bibinfo{person}{Falco~J.
  Bargagli-Stoffi}.} \bibinfo{year}{2026}\natexlab{}.
\newblock \showarticletitle{Fair Policy Learning under Bipartite Network
  Interference: Learning Fair and Cost-Effective Environmental Policies}.
\newblock \bibinfo{journal}{\emph{arXiv preprint arXiv:2601.00531}}
  (\bibinfo{year}{2026}).
\newblock


\bibitem[Kitagawa and Tetenov(2018)]%
        {kitagawa2018who}
\bibfield{author}{\bibinfo{person}{Toru Kitagawa} {and}
  \bibinfo{person}{Aleksey Tetenov}.} \bibinfo{year}{2018}\natexlab{}.
\newblock \showarticletitle{Who Should Be Treated? Empirical Welfare
  Maximization Methods for Treatment Choice}.
\newblock \bibinfo{journal}{\emph{Econometrica}} \bibinfo{volume}{86},
  \bibinfo{number}{2} (\bibinfo{year}{2018}), \bibinfo{pages}{591--616}.
\newblock


\bibitem[Kitagawa and Wang(2023)]%
        {kitagawa2023sequential}
\bibfield{author}{\bibinfo{person}{Toru Kitagawa} {and} \bibinfo{person}{Guanyi
  Wang}.} \bibinfo{year}{2023}\natexlab{}.
\newblock \showarticletitle{Individualized Treatment Allocation in Sequential
  Network Games}.
\newblock \bibinfo{journal}{\emph{arXiv preprint arXiv:2302.05747}}
  (\bibinfo{year}{2023}).
\newblock


\bibitem[Kohavi et~al\mbox{.}(2020)]%
        {kohavi2020trustworthy}
\bibfield{author}{\bibinfo{person}{Ron Kohavi}, \bibinfo{person}{Diane Tang},
  {and} \bibinfo{person}{Ya Xu}.} \bibinfo{year}{2020}\natexlab{}.
\newblock \bibinfo{booktitle}{\emph{Trustworthy Online Controlled Experiments:
  A Practical Guide to A/B Testing}}.
\newblock \bibinfo{publisher}{Cambridge University Press},
  \bibinfo{address}{Cambridge}.
\newblock


\bibitem[Koles{\'a}r and Rothe(2018)]%
        {kolesar2018inference}
\bibfield{author}{\bibinfo{person}{Michal Koles{\'a}r} {and}
  \bibinfo{person}{Christoph Rothe}.} \bibinfo{year}{2018}\natexlab{}.
\newblock \showarticletitle{Inference in Regression Discontinuity Designs with
  a Discrete Running Variable}.
\newblock \bibinfo{journal}{\emph{American Economic Review}}
  \bibinfo{volume}{108}, \bibinfo{number}{8} (\bibinfo{year}{2018}),
  \bibinfo{pages}{2277--2304}.
\newblock


\bibitem[Lee and Shen(2018)]%
        {lee2018winners}
\bibfield{author}{\bibinfo{person}{Minyong~R. Lee} {and} \bibinfo{person}{Milan
  Shen}.} \bibinfo{year}{2018}\natexlab{}.
\newblock \showarticletitle{Winner's Curse: Bias Estimation for Total Effects
  of Features in Online Controlled Experiments}. In
  \bibinfo{booktitle}{\emph{Proceedings of the 24th ACM SIGKDD International
  Conference on Knowledge Discovery \& Data Mining}}.
  \bibinfo{pages}{491--499}.
\newblock


\bibitem[Leung(2022)]%
        {leung2022causal}
\bibfield{author}{\bibinfo{person}{Michael~P. Leung}.}
  \bibinfo{year}{2022}\natexlab{}.
\newblock \showarticletitle{Causal Inference Under Approximate Neighborhood
  Interference}.
\newblock \bibinfo{journal}{\emph{Econometrica}} \bibinfo{volume}{90},
  \bibinfo{number}{1} (\bibinfo{year}{2022}), \bibinfo{pages}{267--293}.
\newblock


\bibitem[Li et~al\mbox{.}(2022)]%
        {li2022interference}
\bibfield{author}{\bibinfo{person}{Hannah Li}, \bibinfo{person}{Geng Zhao},
  \bibinfo{person}{Ramesh Johari}, {and} \bibinfo{person}{Gabriel~Y.
  Weintraub}.} \bibinfo{year}{2022}\natexlab{}.
\newblock \showarticletitle{Interference, Bias, and Variance in Two-Sided
  Marketplace Experimentation: Guidance for Platforms}. In
  \bibinfo{booktitle}{\emph{Proceedings of the ACM Web Conference 2022}}.
  \bibinfo{pages}{182--192}.
\newblock


\bibitem[Li et~al\mbox{.}(2023)]%
        {li2023phased}
\bibfield{author}{\bibinfo{person}{Yufan Li}, \bibinfo{person}{Jialiang Mao},
  {and} \bibinfo{person}{Iavor Bojinov}.} \bibinfo{year}{2023}\natexlab{}.
\newblock \showarticletitle{Balancing Risk and Reward: A Batched-Bandit
  Strategy for Automated Phased Release}. In \bibinfo{booktitle}{\emph{Advances
  in Neural Information Processing Systems 36}}. \bibinfo{pages}{76181--76201}.
\newblock


\bibitem[Lin et~al\mbox{.}(2026)]%
        {lin2026gite}
\bibfield{author}{\bibinfo{person}{Xiaofeng Lin}, \bibinfo{person}{Han Bao},
  {and} \bibinfo{person}{Hisashi Kashima}.} \bibinfo{year}{2026}\natexlab{}.
\newblock \showarticletitle{Treatment Effect Estimation with Differentiated
  Networked Effect on Graph Data}. In \bibinfo{booktitle}{\emph{Proceedings of
  the 32nd ACM SIGKDD Conference on Knowledge Discovery and Data Mining V.2}}.
  \bibinfo{pages}{3116--3127}.
\newblock
\href{https://doi.org/10.1145/3770855.3817678}{doi:\nolinkurl{10.1145/3770855.3817678}}


\bibitem[Loomba and Eckles(2026)]%
        {loomba2026offpolicy}
\bibfield{author}{\bibinfo{person}{Sahil Loomba} {and} \bibinfo{person}{Dean
  Eckles}.} \bibinfo{year}{2026}\natexlab{}.
\newblock \showarticletitle{Off-Policy Causal Estimation in Networks}.
\newblock \bibinfo{journal}{\emph{arXiv preprint arXiv:2609.02756}}
  (\bibinfo{year}{2026}).
\newblock


\bibitem[Lu et~al\mbox{.}(2018)]%
        {lu2018surge}
\bibfield{author}{\bibinfo{person}{Alice Lu}, \bibinfo{person}{Peter~I.
  Frazier}, {and} \bibinfo{person}{Oren Kislev}.}
  \bibinfo{year}{2018}\natexlab{}.
\newblock \showarticletitle{Surge Pricing Moves {Uber}'s Driver-Partners}. In
  \bibinfo{booktitle}{\emph{Proceedings of the 2018 ACM Conference on Economics
  and Computation}}. \bibinfo{pages}{3}.
\newblock


\bibitem[Luo et~al\mbox{.}(2024)]%
        {luo2024policy}
\bibfield{author}{\bibinfo{person}{Shikai Luo}, \bibinfo{person}{Ying Yang},
  \bibinfo{person}{Chengchun Shi}, \bibinfo{person}{Fang Yao},
  \bibinfo{person}{Jieping Ye}, {and} \bibinfo{person}{Hongtu Zhu}.}
  \bibinfo{year}{2024}\natexlab{}.
\newblock \showarticletitle{Policy Evaluation for Temporal and/or Spatial
  Dependent Experiments}.
\newblock \bibinfo{journal}{\emph{Journal of the Royal Statistical Society
  Series B: Statistical Methodology}} \bibinfo{volume}{86}, \bibinfo{number}{3}
  (\bibinfo{year}{2024}), \bibinfo{pages}{623--649}.
\newblock


\bibitem[Ma et~al\mbox{.}(2022)]%
        {ma2022hypergraph}
\bibfield{author}{\bibinfo{person}{Jing Ma}, \bibinfo{person}{Mengting Wan},
  \bibinfo{person}{Longqi Yang}, \bibinfo{person}{Jundong Li},
  \bibinfo{person}{Brent Hecht}, {and} \bibinfo{person}{Jaime Teevan}.}
  \bibinfo{year}{2022}\natexlab{}.
\newblock \showarticletitle{Learning Causal Effects on Hypergraphs}. In
  \bibinfo{booktitle}{\emph{Proceedings of the 28th ACM SIGKDD Conference on
  Knowledge Discovery and Data Mining}}. \bibinfo{pages}{1202--1212}.
\newblock


\bibitem[Montiel~Olea and Plagborg-M{\o}ller(2019)]%
        {montielolea2019simultaneous}
\bibfield{author}{\bibinfo{person}{Jos{\'e}~Luis Montiel~Olea} {and}
  \bibinfo{person}{Mikkel Plagborg-M{\o}ller}.}
  \bibinfo{year}{2019}\natexlab{}.
\newblock \showarticletitle{Simultaneous Confidence Bands: Theory,
  Implementation, and an Application to {SVARs}}.
\newblock \bibinfo{journal}{\emph{Journal of Applied Econometrics}}
  \bibinfo{volume}{34}, \bibinfo{number}{1} (\bibinfo{year}{2019}),
  \bibinfo{pages}{1--17}.
\newblock


\bibitem[Morris et~al\mbox{.}(2019)]%
        {morris2019using}
\bibfield{author}{\bibinfo{person}{Tim~P. Morris}, \bibinfo{person}{Ian~R.
  White}, {and} \bibinfo{person}{Michael~J. Crowther}.}
  \bibinfo{year}{2019}\natexlab{}.
\newblock \showarticletitle{Using Simulation Studies to Evaluate Statistical
  Methods}.
\newblock \bibinfo{journal}{\emph{Statistics in Medicine}}
  \bibinfo{volume}{38}, \bibinfo{number}{11} (\bibinfo{year}{2019}),
  \bibinfo{pages}{2074--2102}.
\newblock


\bibitem[{New York City Taxi and Limousine Commission}(2024)]%
        {nyctlc}
\bibfield{author}{\bibinfo{person}{{New York City Taxi and Limousine
  Commission}}.} \bibinfo{year}{2024}\natexlab{}.
\newblock \bibinfo{title}{{TLC} Trip Record Data}.
\newblock
  \bibinfo{howpublished}{\url{https://www.nyc.gov/site/tlc/about/tlc-trip-record-data.page}}.
\newblock
\shownote{High Volume For-Hire Vehicle trip records, October 2024}.
\newblock


\bibitem[Ni(2025)]%
        {ni2025decision}
\bibfield{author}{\bibinfo{person}{Tu Ni}.} \bibinfo{year}{2025}\natexlab{}.
\newblock \bibinfo{title}{Decision Analytics of Switchback Experiments: A
  Robust Optimization Approach}.
\newblock \bibinfo{howpublished}{SSRN Working Paper 5245482}.
\newblock


\bibitem[Noack and Rothe(2024)]%
        {noack2024bias}
\bibfield{author}{\bibinfo{person}{Claudia Noack} {and}
  \bibinfo{person}{Christoph Rothe}.} \bibinfo{year}{2024}\natexlab{}.
\newblock \showarticletitle{Bias-Aware Inference in Fuzzy Regression
  Discontinuity Designs}.
\newblock \bibinfo{journal}{\emph{Econometrica}} \bibinfo{volume}{92},
  \bibinfo{number}{3} (\bibinfo{year}{2024}), \bibinfo{pages}{687--711}.
\newblock
\href{https://doi.org/10.3982/ECTA19466}{doi:\nolinkurl{10.3982/ECTA19466}}


\bibitem[Prinster et~al\mbox{.}(2026)]%
        {prinster2026cpc}
\bibfield{author}{\bibinfo{person}{Drew Prinster}, \bibinfo{person}{Clara
  Fannjiang}, \bibinfo{person}{Ji~Won Park}, \bibinfo{person}{Kyunghyun Cho},
  \bibinfo{person}{Anqi Liu}, \bibinfo{person}{Suchi Saria}, {and}
  \bibinfo{person}{Samuel Stanton}.} \bibinfo{year}{2026}\natexlab{}.
\newblock \showarticletitle{Conformal Policy Control}. In
  \bibinfo{booktitle}{\emph{Proceedings of the 43rd International Conference on
  Machine Learning}}.
\newblock
\shownote{arXiv:2603.02196}.
\newblock


\bibitem[Rinaldo et~al\mbox{.}(2019)]%
        {rinaldo2019bootstrapping}
\bibfield{author}{\bibinfo{person}{Alessandro Rinaldo}, \bibinfo{person}{Larry
  Wasserman}, {and} \bibinfo{person}{Max G'Sell}.}
  \bibinfo{year}{2019}\natexlab{}.
\newblock \showarticletitle{Bootstrapping and Sample Splitting for
  High-Dimensional, Assumption-Lean Inference}.
\newblock \bibinfo{journal}{\emph{The Annals of Statistics}}
  \bibinfo{volume}{47}, \bibinfo{number}{6} (\bibinfo{year}{2019}),
  \bibinfo{pages}{3438--3469}.
\newblock


\bibitem[Romano and Wolf(2005)]%
        {romano2005stepwise}
\bibfield{author}{\bibinfo{person}{Joseph~P. Romano} {and}
  \bibinfo{person}{Michael Wolf}.} \bibinfo{year}{2005}\natexlab{}.
\newblock \showarticletitle{Stepwise Multiple Testing as Formalized Data
  Snooping}.
\newblock \bibinfo{journal}{\emph{Econometrica}} \bibinfo{volume}{73},
  \bibinfo{number}{4} (\bibinfo{year}{2005}), \bibinfo{pages}{1237--1282}.
\newblock


\bibitem[Sahoo and Wager(2025)]%
        {sahoo2025competing}
\bibfield{author}{\bibinfo{person}{Roshni Sahoo} {and} \bibinfo{person}{Stefan
  Wager}.} \bibinfo{year}{2025}\natexlab{}.
\newblock \showarticletitle{Policy Learning with Competing Agents}.
\newblock \bibinfo{journal}{\emph{Operations Research}} \bibinfo{volume}{73},
  \bibinfo{number}{5} (\bibinfo{year}{2025}), \bibinfo{pages}{2515--2529}.
\newblock


\bibitem[S{\"a}vje(2024)]%
        {savje2024causal}
\bibfield{author}{\bibinfo{person}{Fredrik S{\"a}vje}.}
  \bibinfo{year}{2024}\natexlab{}.
\newblock \showarticletitle{Causal Inference with Misspecified Exposure
  Mappings: Separating Definitions and Assumptions}.
\newblock \bibinfo{journal}{\emph{Biometrika}} \bibinfo{volume}{111},
  \bibinfo{number}{1} (\bibinfo{year}{2024}), \bibinfo{pages}{1--15}.
\newblock


\bibitem[Sawarni et~al\mbox{.}(2026)]%
        {sawarni2025abstention}
\bibfield{author}{\bibinfo{person}{Ayush Sawarni}, \bibinfo{person}{Jikai Jin},
  \bibinfo{person}{Justin Whitehouse}, {and} \bibinfo{person}{Vasilis
  Syrgkanis}.} \bibinfo{year}{2026}\natexlab{}.
\newblock \showarticletitle{Policy Learning with Abstention}. In
  \bibinfo{booktitle}{\emph{Proceedings of the 29th International Conference on
  Artificial Intelligence and Statistics}} \emph{(\bibinfo{series}{Proceedings
  of Machine Learning Research}, Vol.~\bibinfo{volume}{300})}.
  \bibinfo{pages}{4465--4473}.
\newblock


\bibitem[Schr{\"o}der et~al\mbox{.}(2026)]%
        {schroder2026misspecified}
\bibfield{author}{\bibinfo{person}{Maresa Schr{\"o}der},
  \bibinfo{person}{Miruna Oprescu}, \bibinfo{person}{Stefan Feuerriegel}, {and}
  \bibinfo{person}{Nathan Kallus}.} \bibinfo{year}{2026}\natexlab{}.
\newblock \bibinfo{title}{Causal Inference on Networks under Misspecified
  Exposure Mappings: A Partial Identification Framework}.
\newblock
\shownote{arXiv:2602.03459}.
\newblock


\bibitem[Schultzberg et~al\mbox{.}(2026)]%
        {schultzberg2026risk}
\bibfield{author}{\bibinfo{person}{M{\aa}rten Schultzberg},
  \bibinfo{person}{Sebastian Ankargren}, {and} \bibinfo{person}{Mattias
  Fr{\aa}nberg}.} \bibinfo{year}{2026}\natexlab{}.
\newblock \showarticletitle{Risk-Aware Product Decisions in {A/B} Tests with
  Multiple Metrics}.
\newblock \bibinfo{journal}{\emph{Journal of Statistical Planning and
  Inference}}  \bibinfo{volume}{245} (\bibinfo{year}{2026}),
  \bibinfo{pages}{106393}.
\newblock


\bibitem[Shekhar and Howard(2026a)]%
        {shekhar2026designs}
\bibfield{author}{\bibinfo{person}{Prashant Shekhar} {and}
  \bibinfo{person}{Caroline Howard}.} \bibinfo{year}{2026}\natexlab{a}.
\newblock \showarticletitle{Choosing Online Experiment Designs under
  Interference in Ads, Recommendations, and Member-Experience Systems}.
\newblock \bibinfo{journal}{\emph{arXiv preprint arXiv:2605.25290}}
  (\bibinfo{year}{2026}).
\newblock


\bibitem[Shekhar and Howard(2026b)]%
        {shekhar2026launch}
\bibfield{author}{\bibinfo{person}{Prashant Shekhar} {and}
  \bibinfo{person}{Caroline Howard}.} \bibinfo{year}{2026}\natexlab{b}.
\newblock \showarticletitle{Decision Support for Marketplace Policies under
  Incomplete Evidence: From Replay to Launch Readiness}.
\newblock \bibinfo{journal}{\emph{arXiv preprint arXiv:2605.12840}}
  (\bibinfo{year}{2026}).
\newblock


\bibitem[Shekhar and Howard(2026c)]%
        {shekhar2026support}
\bibfield{author}{\bibinfo{person}{Prashant Shekhar} {and}
  \bibinfo{person}{Caroline Howard}.} \bibinfo{year}{2026}\natexlab{c}.
\newblock \showarticletitle{Support-Aware Offline Policy Selection for
  Advertising Marketplaces}.
\newblock \bibinfo{journal}{\emph{arXiv preprint arXiv:2605.21736}}
  (\bibinfo{year}{2026}).
\newblock


\bibitem[Shi et~al\mbox{.}(2023)]%
        {shi2023multiagent}
\bibfield{author}{\bibinfo{person}{Chengchun Shi}, \bibinfo{person}{Runzhe
  Wan}, \bibinfo{person}{Ge Song}, \bibinfo{person}{Shikai Luo},
  \bibinfo{person}{Hongtu Zhu}, {and} \bibinfo{person}{Rui Song}.}
  \bibinfo{year}{2023}\natexlab{}.
\newblock \showarticletitle{A Multiagent Reinforcement Learning Framework for
  Off-Policy Evaluation in Two-Sided Markets}.
\newblock \bibinfo{journal}{\emph{The Annals of Applied Statistics}}
  \bibinfo{volume}{17}, \bibinfo{number}{4} (\bibinfo{year}{2023}),
  \bibinfo{pages}{2701--2722}.
\newblock


\bibitem[Shirani et~al\mbox{.}(2025)]%
        {shirani2025validate}
\bibfield{author}{\bibinfo{person}{Sadegh Shirani}, \bibinfo{person}{Yuwei
  Luo}, \bibinfo{person}{William Overman}, \bibinfo{person}{Ruoxuan Xiong},
  {and} \bibinfo{person}{Mohsen Bayati}.} \bibinfo{year}{2025}\natexlab{}.
\newblock \showarticletitle{Can We Validate Counterfactual Estimations in the
  Presence of General Network Interference?}
\newblock \bibinfo{journal}{\emph{arXiv preprint arXiv:2502.01106}}
  (\bibinfo{year}{2025}).
\newblock


\bibitem[Thomas et~al\mbox{.}(2019)]%
        {thomas2019seldonian}
\bibfield{author}{\bibinfo{person}{Philip~S. Thomas}, \bibinfo{person}{Bruno
  Castro~da Silva}, \bibinfo{person}{Andrew~G. Barto}, \bibinfo{person}{Stephen
  Giguere}, \bibinfo{person}{Yuriy Brun}, {and} \bibinfo{person}{Emma
  Brunskill}.} \bibinfo{year}{2019}\natexlab{}.
\newblock \showarticletitle{Preventing Undesirable Behavior of Intelligent
  Machines}.
\newblock \bibinfo{journal}{\emph{Science}} \bibinfo{volume}{366},
  \bibinfo{number}{6468} (\bibinfo{year}{2019}), \bibinfo{pages}{999--1004}.
\newblock


\bibitem[Thomas et~al\mbox{.}(2015a)]%
        {thomas2015hcope}
\bibfield{author}{\bibinfo{person}{Philip~S. Thomas}, \bibinfo{person}{Georgios
  Theocharous}, {and} \bibinfo{person}{Mohammad Ghavamzadeh}.}
  \bibinfo{year}{2015}\natexlab{a}.
\newblock \showarticletitle{High-Confidence Off-Policy Evaluation}. In
  \bibinfo{booktitle}{\emph{Proceedings of the 29th AAAI Conference on
  Artificial Intelligence}}. \bibinfo{pages}{3000--3006}.
\newblock


\bibitem[Thomas et~al\mbox{.}(2015b)]%
        {thomas2015hcpi}
\bibfield{author}{\bibinfo{person}{Philip~S. Thomas}, \bibinfo{person}{Georgios
  Theocharous}, {and} \bibinfo{person}{Mohammad Ghavamzadeh}.}
  \bibinfo{year}{2015}\natexlab{b}.
\newblock \showarticletitle{High Confidence Policy Improvement}. In
  \bibinfo{booktitle}{\emph{Proceedings of the 32nd International Conference on
  Machine Learning}} \emph{(\bibinfo{series}{PMLR},
  Vol.~\bibinfo{volume}{37})}. \bibinfo{pages}{2380--2388}.
\newblock


\bibitem[Ugander et~al\mbox{.}(2013)]%
        {ugander2013graph}
\bibfield{author}{\bibinfo{person}{Johan Ugander}, \bibinfo{person}{Brian
  Karrer}, \bibinfo{person}{Lars Backstrom}, {and} \bibinfo{person}{Jon
  Kleinberg}.} \bibinfo{year}{2013}\natexlab{}.
\newblock \showarticletitle{Graph Cluster Randomization: Network Exposure to
  Multiple Universes}. In \bibinfo{booktitle}{\emph{Proceedings of the 19th ACM
  SIGKDD International Conference on Knowledge Discovery and Data Mining}}.
  \bibinfo{pages}{329--337}.
\newblock


\bibitem[van~den Akker(2025)]%
        {vandenakker2025profit}
\bibfield{author}{\bibinfo{person}{Bram van~den Akker}.}
  \bibinfo{year}{2025}\natexlab{}.
\newblock \showarticletitle{Direct Profit Estimation Using Uplift Modeling
  under Clustered Network Interference}.
\newblock \bibinfo{journal}{\emph{arXiv preprint arXiv:2509.01558}}
  (\bibinfo{year}{2025}).
\newblock


\bibitem[Viviano(2025)]%
        {viviano2025targeting}
\bibfield{author}{\bibinfo{person}{Davide Viviano}.}
  \bibinfo{year}{2025}\natexlab{}.
\newblock \showarticletitle{Policy Targeting under Network Interference}.
\newblock \bibinfo{journal}{\emph{Review of Economic Studies}}
  \bibinfo{volume}{92}, \bibinfo{number}{2} (\bibinfo{year}{2025}),
  \bibinfo{pages}{1257--1292}.
\newblock
\href{https://doi.org/10.1093/restud/rdae041}{doi:\nolinkurl{10.1093/restud/rdae041}}


\bibitem[Wager and Xu(2021)]%
        {wager2021experimenting}
\bibfield{author}{\bibinfo{person}{Stefan Wager} {and} \bibinfo{person}{Kuang
  Xu}.} \bibinfo{year}{2021}\natexlab{}.
\newblock \showarticletitle{Experimenting in Equilibrium}.
\newblock \bibinfo{journal}{\emph{Management Science}} \bibinfo{volume}{67},
  \bibinfo{number}{11} (\bibinfo{year}{2021}), \bibinfo{pages}{6694--6715}.
\newblock


\bibitem[Westfall and Young(1993)]%
        {westfall1993resampling}
\bibfield{author}{\bibinfo{person}{Peter~H. Westfall} {and}
  \bibinfo{person}{S.~Stanley Young}.} \bibinfo{year}{1993}\natexlab{}.
\newblock \bibinfo{booktitle}{\emph{Resampling-Based Multiple Testing: Examples
  and Methods for p-Value Adjustment}}.
\newblock \bibinfo{publisher}{Wiley}, \bibinfo{address}{New York}.
\newblock


\bibitem[White(2000)]%
        {white2000reality}
\bibfield{author}{\bibinfo{person}{Halbert White}.}
  \bibinfo{year}{2000}\natexlab{}.
\newblock \showarticletitle{A Reality Check for Data Snooping}.
\newblock \bibinfo{journal}{\emph{Econometrica}} \bibinfo{volume}{68},
  \bibinfo{number}{5} (\bibinfo{year}{2000}), \bibinfo{pages}{1097--1126}.
\newblock


\bibitem[Wu et~al\mbox{.}(2026)]%
        {wu2026switchback}
\bibfield{author}{\bibinfo{person}{Yifan Wu}, \bibinfo{person}{Ramesh Johari},
  \bibinfo{person}{Vasilis Syrgkanis}, {and} \bibinfo{person}{Gabriel~Y.
  Weintraub}.} \bibinfo{year}{2026}\natexlab{}.
\newblock \showarticletitle{Switchback Price Experiments with Forward-Looking
  Demand}. In \bibinfo{booktitle}{\emph{Proceedings of the 27th ACM Conference
  on Economics and Computation}}.
\newblock
\shownote{arXiv:2410.14904}.
\newblock


\bibitem[Wu and Yan(2026)]%
        {wu2026marketplace}
\bibfield{author}{\bibinfo{person}{Yufei Wu} {and} \bibinfo{person}{Zhen Yan}.}
  \bibinfo{year}{2026}\natexlab{}.
\newblock \showarticletitle{Personalizing Marketplace Policies with Competing
  Objectives and Constrained Experiments: Evidence from a Job Marketplace}. In
  \bibinfo{booktitle}{\emph{Proceedings of the 32nd ACM SIGKDD Conference on
  Knowledge Discovery and Data Mining V.2}}. \bibinfo{pages}{8279--8288}.
\newblock
\href{https://doi.org/10.1145/3770855.3818460}{doi:\nolinkurl{10.1145/3770855.3818460}}


\end{thebibliography}
\appendix
\section{Proofs}
\label{app:proofs}

\subsection{Proof of Theorem~\ref{thm:band}}
\label{app:biasaware}

The argument conditions on the design, the own treatments and exposures of every calibration round, which alone determine the bins. Conditionally on the design, the bins and weights below are fixed, and the statement holds for every binning outcome, as in bias-aware inference with a discrete running variable \cite{kolesar2018inference,imbens2019optimized}.

\paragraph{Estimation weights}
Fix a unit $i$, an own treatment $t$, and an exposure $e$ of interest, and let $S$ be the bin of $(i, t)$ assigned to $e$, with $n_S$ rounds, mean exposure $\bar{e}_S$, and $Q_S = \sum_{r \in S} (e^r_i - \bar{e}_S)^2$. The line fitted in $S$ and evaluated at $e$ is $\hat{\mu}_i(t, e) = \sum_{r \in S} w_r(e)\, y^r_i$ with
\begin{equation}
w_r(e) = \frac{1}{n_S} + \frac{(e - \bar{e}_S)(e^r_i - \bar{e}_S)}{Q_S},
\label{eq:weights}
\end{equation}
and a bin that contributes its mean has $w_r(e) = 1/n_S$. These weights satisfy $\sum_r w_r(e) = 1$ and, for the line, $\sum_r w_r(e)\, e^r_i = e$ for every $e$, inside or outside the recorded range. Substituting these weights into \eqref{eq:strat}, every held-out estimate is a linear combination of the calibration outcomes, $\gcal(a) = \sum_r \sum_i c^r_i(a)\, y^r_i$, with coefficients $c^r_i(a)$ that depend on the design and the candidate only.

\paragraph{Bias bound}
For an evaluation point $(t, e)$ in a bin $S$ fitted by a line the bias bound is \eqref{eq:biasbound}. In a bin that contributes its mean it is $b_i(t, e) = L_i |\bar{e}_S - e| + (M_i / 2n_S) \sum_{r \in S} (e^r_i - e)^2$, with $|\partial_e \mu_i(t, \cdot)| \le L_i$. A unit with no round at own treatment $t$ has $b_i(t, e) = \infty$, so its candidates do not pass. \eqref{eq:biasbound} is the exact worst-case bias of the fitted line over the class of responses whose second-order Taylor remainder at $e$ is bounded by $M_i/2$ times the squared distance \citep[eq.~(11)]{armstrong2018optimal}, and the curvature class of Theorem~\ref{thm:band} is contained in that class. $M_i$ and $L_i$ come from a quadratic fit in exposure on the training half, per unit and own treatment, as $2|b_2|$ and $\max(|b_1|, |b_1 + 2 b_2|)$, a global-polynomial rule of thumb \cite{armstrong2020simple,imbens2019optimized}. A unit with fewer than six training rounds or three distinct exposures at that own treatment takes the median of the values fitted for the other units in $P$ at the same own treatment.

\paragraph{Round-level scores}
The residual $\hat{\varepsilon}^r_i$ of round $r$ from the fit of its bin is scaled by $(1 - h^r_i)^{-1/2}$, with $h^r_i$ the round's leverage in that fit. A round with leverage one, such as the only round at one of two exposure values in a bin, has a zero residual. Its term in $u^r(a)$ is replaced in every draw by $c^r_i(a)\,\hat{s}_{it}\,\zeta$, with $\zeta$ an independent standard normal and $\hat{s}_{it}$ the residual standard deviation of unit $i$ at own treatment $t$ pooled over its bins, and $\hat{\nu}(a)^2$ adds $c^r_i(a)^2 \hat{s}_{it}^2$. The term is drawn afresh in every multiplier draw, independent across units and rounds, and shared by all candidates within a draw. Such rounds vanish as counts grow. A candidate with $\hat{\nu}(a) = 0$ is left out of the maximum in \eqref{eq:band}, and its bound is $\gcal(a) - \allow(a)$.

\paragraph{Asymptotic regime}
The graph, $P$, the training half, $\cand$, and the bounds $M_i$, $L_i$ are fixed. Every unit is treated with probability $\rho \in (0, 1)$, and the exposure support is finite because the graph is. Both $R_{\mathrm{cal}}$ and $B_{\mathrm{boot}}$ tend to infinity. Write $g^r(a) = \sum_i c^r_i(a)\, y^r_i$ for the contribution of round $r$ and $V(a) = \sum_r \mathrm{Var}(g^r(a) \mid \text{design})$. The conditions of Theorem~\ref{thm:band} are $\mathbb{E}[y^r_i \mid \text{design}] = \mu_i(a^r_i, e^r_i)$, the conditional covariance matrix of each round's outcomes over $P$ with eigenvalues in $[\sigma_{\min}^2, \sigma_{\max}^2]$, $\sigma_{\min} > 0$, which allows any non-degenerate dependence among units within a round, $\mathbb{E}[(g^r(a) - \mathbb{E} g^r(a))^4 \mid \text{design}] \le C\, \mathrm{Var}(g^r(a) \mid \text{design})^2$ (bounded kurtosis), and no dominance, $\max_r \mathrm{Var}(g^r(a) \mid \text{design}) / V(a) \to 0$ for every $a \in \cand$.

\paragraph{Proof of Theorem~\ref{thm:band}}
(i) Fix an evaluation point in a bin fitted by a line. Write $\mu_i(t, e^r_i) = \mu_i(t, e) + \partial_e \mu_i(t, e)(e^r_i - e) + \varrho_r$ with $|\varrho_r| \le (M_i/2)(e^r_i - e)^2$. Because the weights reproduce constants and the identity, $\mathbb{E}[\hat{\mu}_i(t, e) \mid \text{design}] - \mu_i(t, e) = \sum_r w_r(e)\, \varrho_r$, bounded in absolute value by \eqref{eq:biasbound}, and choosing $\varrho_r$ with the sign of $w_r(e)$ attains it over the Taylor class. For an evaluation point in a bin fitted by its mean, the first-order term survives as $\partial_e \mu_i(t, e)(\bar{e}_S - e)$, bounded by $L_i |\bar{e}_S - e|$. The bias of $\hat{\mu}_i(a_i, e_i(a)) - \hat{\mu}_i(0, 0)$ is at most the sum of the two bounds, and summing over $P$ gives $\allow(a)$. The pair of curves $\mu_i(1, \cdot)$ and $\mu_i(0, \cdot)$ carries no cross-restriction, so for $a_i = 1$ the sum is attained, the argument of \citet[Sections~2 and 3]{noack2024bias} for a difference of two fitted values.

(ii) Conditionally on the design, $\gcal(a) - \mathbb{E}[\gcal(a) \mid \text{design}] = \sum_r \sum_i c^r_i(a)(y^r_i - \mu_i(a^r_i, e^r_i))$ is a sum over independent rounds with fixed coefficients. By bounded kurtosis, $\sum_r \mathbb{E}[(g^r(a) - \mathbb{E} g^r(a))^4 \mid \text{design}] / V(a)^2 \le C \max_r \mathrm{Var}(g^r(a) \mid \text{design}) / V(a) \to 0$, Lyapunov's condition, so the vector over $\cand$, each coordinate divided by its conditional standard deviation, converges to a centered Gaussian vector $Z$ with unit variances along any subsequence on which its correlation matrix converges, which suffices for the limit of the probability below. Lemma~\ref{lem:sandwich} gives $\hat{\nu}(a) / \mathrm{sd}(\gcal(a) \mid \text{design}) \to 1$ and the consistency of $c_\alpha$ for the $(1 - \alpha)$ quantile of $\max_a Z_a$, so by Slutsky's theorem $\Pr[\max_a (\gcal(a) - \mathbb{E}[\gcal(a) \mid \text{design}]) / \hat{\nu}(a) > c_\alpha \mid \text{design}] \to \alpha$. On the complement, $\gcal(a) - c_\alpha \hat{\nu}(a) \le \mathbb{E}[\gcal(a) \mid \text{design}] \le \gtrue(a) + \allow(a)$ for every $a$ by (i), which is the event that every bound holds. A violation requires $\gtrue(\ahat) < -\slack \le \ell_\alpha(\ahat)$, which lies in the complement of that event, whichever $\ahat$ is selected. \qed

\begin{lemma}[Sandwich variance and multiplier quantile]
\label{lem:sandwich}
Under the assumptions of Theorem~\ref{thm:band}, $\hat{\nu}(a)^2 / \mathrm{Var}(\gcal(a) \mid \text{design}) \to 1$ in probability for every $a \in \cand$ with $\mathrm{Var}(\gcal(a) \mid \text{design}) > 0$, and, given the data, the multiplier vector with coordinates $\sum_r \xi^r_b\, u^r(a) / \hat{\nu}(a)$ converges in law to $Z$, so $c_\alpha$ converges in probability to the $(1 - \alpha)$ quantile of $\max_a Z_a$ as $B_{\mathrm{boot}} \to \infty$.
\end{lemma}

\paragraph{Proof}
Within a bin, the residuals combine noise with the lack of fit of the line or bin mean to the conditional mean, and the lack-of-fit component is zero in a single-value bin. As $R_{\mathrm{cal}}$ grows every exposure value of positive probability reaches $n_{\min}$ rounds, so the bins become single-value with probability tending to one and the lack-of-fit term vanishes. In a single-value bin the fit is the bin mean and the leverages are $1/n_S \to 0$, so the leverage-scaled residual of unit $i$ in round $r$ is its noise plus a term $d^r_i$, made of the bin's average noise and the scaling correction, with $\mathbb{E}[(d^r_i)^2 \mid \text{design}] \le C \sigma^2_{\max} / n_S$. Write $\tilde{g}^r(a) = g^r(a) - \mathbb{E}[g^r(a) \mid \text{design}]$ and $D^r(a) = u^r(a) - \tilde{g}^r(a) = \sum_{i \in P} c^r_i(a)\, d^r_i$. By the Cauchy-Schwarz inequality, $D^r(a)^2 \le |P| \sum_{i \in P} c^r_i(a)^2 (d^r_i)^2$, and because the eigenvalues of each round's outcome covariance over $P$ are at least $\sigma^2_{\min}$, whatever the dependence among units, $\mathrm{Var}(g^r(a) \mid \text{design}) \ge \sigma^2_{\min} \sum_{i \in P} c^r_i(a)^2$. Hence $\mathbb{E}[\sum_r D^r(a)^2 \mid \text{design}] / V(a)$ is at most $|P|\, C (\sigma^2_{\max} / \sigma^2_{\min}) \max_S 1/n_S$, which tends to zero with $|P|$ fixed on the fixed graph. The sum $\sum_r \tilde{g}^r(a)^2 / V(a)$ has mean one and, by independence across rounds and bounded kurtosis, variance at most $C \max_r \mathrm{Var}(g^r(a) \mid \text{design}) / V(a) \to 0$ by no dominance, so it converges to one. With the Cauchy-Schwarz inequality on the cross term, $\hat{\nu}(a)^2 / V(a)$, which is $\sum_r (\tilde{g}^r(a) + D^r(a))^2 / V(a)$, tends to one in probability. Given the data, $\sum_r \xi^r_b\, u^r(a)$ is Gaussian with covariance $\sum_r u^r(a) u^r(a')$. Divided by $\hat{\nu}(a) \hat{\nu}(a')$, this covariance converges to the correlation of $Z_a$ and $Z_{a'}$ by the same argument, so the multiplier vector converges to $Z$. The maximum is continuous and the limit has a continuous distribution function, strictly increasing on its support, which gives $c_\alpha$, as for the multiplier bootstrap of maxima \citep[Theorems~3.1 and 3.2]{chernozhukov2013gaussian}. \qed

\paragraph{Extension to other linear smoothers}
The proof of Theorem~\ref{thm:band} uses the within-bin line only through its weights. Let a fit give $\hat{\mu}_i(t, e) = \sum_r w_r(e)\, y^r_i$ over the calibration rounds of unit $i$ at own treatment $t$, with weights fixed given the design and the training half and $\sum_r w_r(e) = 1$, and set
\begin{equation}
b_i(t, e) = L_i \Big| \sum_r w_r(e)\, e^r_i - e \Big| + \frac{M_i}{2} \sum_r |w_r(e)|\, (e^r_i - e)^2,
\label{eq:allowsmoother}
\end{equation}
with $b_i(t, e) = \infty$ where the fit defines no weights. The Taylor expansion gives the bias $\partial_e \mu_i(t, e) \cdot ( \sum_r w_r(e)\, e^r_i - e ) + \sum_r w_r(e)\, \varrho_r$, which \eqref{eq:allowsmoother} bounds, so part (i) of Theorem~\ref{thm:band} holds, with the scores formed from the residuals and leverages of the same fit, for the within-bin line, the kernel local-linear fit, and the coarsened H\'ajek estimate. For part (ii) the coefficients are fixed given the design, and Lemma~\ref{lem:sandwich} needs residuals that converge to the noise. Those of the within-bin line do so as its bins become single-value, and those of the coarsened estimate are deviations from exact-exposure means. The kernel fit keeps its bandwidth, so part (ii) requires in addition that the smoothing error in its residuals vanish. For the within-bin line, \eqref{eq:weights} reproduces lines, so the first term vanishes and \eqref{eq:allowsmoother} is \eqref{eq:biasbound}, and for a mean bin it is the mean-bin bias bound above. The kernel fit is weighted least squares in $(1, e^r_i - e)$ with a bandwidth fixed by the training half and the design. Its normal equations give $\sum_r w_r(e) = 1$ and $\sum_r w_r(e)(e^r_i - e) = 0$, so it reproduces lines. The coarsened H\'ajek estimate puts weight $1/n'$ on each of the $n'$ rounds at $e'$, the nearest recorded exposure at or below $e$, or the lowest recorded exposure when $e$ lies below all of them. It reproduces constants but not lines, and \eqref{eq:allowsmoother} becomes $L_i |e - e'| + (M_i/2)(e - e')^2$, zero at a recorded evaluation point.

\subsection{Coverage failure and unsafe selection}
\label{app:events}

Every method returns an element of $\cand \cup \{\emptyset\}$, where the status quo $\emptyset$ never violates the guardrail, as in safe policy improvement and risk control \cite{thomas2015hcpi,thomas2019seldonian,angelopoulos2025learn,cho2025cspimt}, and we treat $\emptyset$ as a candidate with gain 0. For a method with lower bounds $\ell(a)$ and pass set $\{a : \ell(a) \ge -\slack\}$, let $E_1$ be the event that some candidate has $\gtrue(a) < \ell(a)$, $E_2$ the event that some unsafe candidate passes, and $E_3$ the event that the returned allocation is unsafe.

\begin{lemma}[Nested error events]
\label{lem:events}
The events satisfy $E_3 \subseteq E_2 \subseteq E_1$. Hence $\Pr[E_3] \le \Pr[E_2] \le \Pr[E_1]$, and a band with simultaneous coverage $1 - \alpha$ bounds all three by $\alpha$, whichever passing candidate is returned and from whichever data its gain is ranked.
\end{lemma}

\paragraph{Proof}
If an unsafe candidate passes then $\gtrue(a) < -\slack \le \ell(a)$, so $E_2 \subseteq E_1$. The returned allocation is either $\emptyset$, which is safe, or a passing candidate, so $E_3 \subseteq E_2$. \qed

Theorem~\ref{thm:band} bounds $E_1$, the fixed-sequence test controls $E_2$ without one \cite{angelopoulos2025learn}, and the per-candidate band controls neither.

\section{Additional experimental details and results}
\label{app:nyc}

\paragraph{Selection rules and implementation}
Unless stated, a method returns the passing candidate with the highest $\wcal$, or the status quo when none has $\wcal > 0$. The subscript tr marks training-half estimates, $\hat{\sigma}_{\mathrm{tr}}(a)$ is the standard deviation of $\gcal_{\mathrm{tr}}(a)$ over 1,000 resamples of the training rounds with replacement, $z_{1-\alpha}$ is the standard normal quantile, and $\eta$ scales the predicted margin. \MARGIN-$m$ ($m \in \{1, 1.5, 2, 3, 5\}$) ranks by $\what$. The fixed-sequence test \cite{angelopoulos2025learn} tests candidates in decreasing order of $(\gcal_{\mathrm{tr}}(a) + \slack) / \hat{\sigma}_{\mathrm{tr}}(a)$, each with its own bound, until the first failure. The Bonferroni band uses level $\alpha / |\cand|$ per candidate, and the Seldonian procedure \cite{thomas2019seldonian} picks on the training half the highest-gain candidate whose predicted bound $\gcal_{\mathrm{tr}}(a) - \eta z_{1-\alpha} \hat{\sigma}_{\mathrm{tr}}(a)$ meets the tolerance, and deploys it if its calibration-half band over that single candidate, without a bias bound, meets the tolerance.

Our adaptation of CSPI-MT's screen \cite[Algorithm~4]{cho2025cspimt} first picks on the training half the candidate with the largest $(\gcal_{\mathrm{tr}}(a) + \slack) / \hat{\sigma}_{\mathrm{tr}}(a)$. Candidates with at least its training gain are ordered by decreasing absolute correlation with it. Each is added only if every training-half bound of the resulting set stays at or above $-\slack$. The calibration band covers the screened set only. Relative to the original, the adaptation uses random halves rather than a 20/80 split, held-out gain rather than a whole-log re-estimate, and separate guardrail and gain quantities rather than one improvement. Screening uses no bias bound, whereas the calibration band uses the reported $k$.

Unless stated, the results below use the default band and bias bound ($k = 1$), with the status quo as a candidate (Appendix~\ref{app:events}). With $n$ selections, the binomial test of Section~\ref{sec:experiments} rejects at a count $r$ with $r - 1 \ge 0.05 n$, and the largest count whose Clopper-Pearson bound lies below 0.05 is at most $0.05 n - 1$, for every $n$ in our tables, the conditions under which \citet[Theorem~4]{hoeffding1956distribution} preserves the level when the trial probabilities differ across settings.

\paragraph{All-zone and per-setting results}
In the all-zone scenario the bias bound raises coverage from 0.92--0.96 to 0.97--0.99 at every $R$ of Table~\ref{tab:allow} and abstention by at most 0.13. No selection violates, and at $R = 480$ the gain halves. At $R = 480$ and $960$, the worst of the 24 settings has 1 violating selection in 200 replications (pointwise upper bound 0.023), underserved-only, and none all-zone.

\paragraph{Sensitivity to experimental settings}
With $n_{\min} = 30$, abstention is 0.35 to 0.68 at $R \le 960$. At 20 percent of the edges rewired, abstention is 0.57. Rounds that randomize contiguous clusters of about eight zones record mostly extreme exposures: without the bias bound the band covers in 0.84 to 1.00, and with it \METHOD{} abstains in 0.94 to 1.00.

\paragraph{Exact-exposure and kernel estimates}
H\'ajek-exact and Horvitz--Thompson leave an evaluation point at an unrecorded exposure undefined, so a candidate that needs one does not pass. At $R = 960$, H\'ajek-exact and Horvitz--Thompson define 17 and 19 percent of the candidates. Horvitz--Thompson needs one recorded round at each evaluation point, whereas H\'ajek-exact needs two, one for a residual. Both take their scores from residuals around the exact-exposure means. Horvitz--Thompson weights a round by $1/(R_{\mathrm{cal}}\, \pi)$, $\pi$ the design probability of the evaluation point. The kernel fit widens its window until it holds three rounds with two distinct exposures. Fewer than 0.2 percent of evaluation points, and 3 percent of candidates at $R = 960$ (4 at $R = 480$), stay undefined.

\paragraph{Exposure mappings and response shapes}
We vary the loss term in \eqref{eq:dgp} while the default estimator retains $e_i(a)$. Demand-weighted exposure is $\sum_j A_{ij}a_jd_j/\sum_j A_{ij}d_j$. Two-hop exposure is $0.75\,e_i(a)+0.25\,e_i^{(2)}(a)$, where $e_i^{(2)}$ is the treated share at graph distance two. Within two hops, distance-decay weights are $\exp(-\mathrm{dist}_{ij}/l)$, $l$ the median edge length. Trip-flow exposure is $\sum_j W_{ij}a_j$, where $W$ is the row-normalized matrix of symmetrized trip counts between distinct zones, which the flow-mapping estimator uses in place of the one-hop share.

Carryover replaces exposure in round $r$ with $(1-c)e_i(a^r)+c\,e_i(a^{r-1})$, starting at $a^0=0$. For deployment truth, both the chosen allocation and the status quo are held across consecutive rounds. Their current and previous exposures therefore coincide, leaving $\gtrue(a)$ unchanged from the no-carryover model. The lagged mapping keeps the bins by own treatment and current exposure, adds the previous round's own treatment and exposure as regressors of the within-bin fit, computes the bias bound from that fit's weights, and evaluates the lag regressors at the held allocation's values. At $R=3840$, underserved-only unconditional violations are 0.138 under two-hop spillovers and 0.083 under carryover with $c=0.25$.

Section~\ref{sec:biasbound} reports the convex response $h(e)=e^2$ at $\gamma\in\{4,6\}$. Under a logistic threshold at $e_0\in\{0.3,0.5\}$, no selection violates and coverage is 0.97 to 0.99.

\paragraph{Residual check for omitted spillovers}
The check uses 500 permutation draws per covariate and rejects when either $p$ is below 0.025. At $R = 960$ it rejects in 0.048 of replications where the mapping holds, 0.55 under two-hop spillover, 0.75 under distance decay, 0.36 under trip flows, and 0.64 and 0.84 under carryover.

\paragraph{Contagion testbed}
We simulate epidemics with EoN on a 200-unit random geometric graph at two densities. Gain is the never-infected count. Each allocation's true effect is approximated by 2,000 Monte Carlo epidemics. Immunizing a set moves the three index cases toward the rest of the graph, so the vulnerable group's infections (top risk-mass quintile) can rise, and the guardrail caps them at 0.25 or 0.5 times their untreated level. Transmission, budget, treatment rate, and the surface vary over 1,280 replications at 300 and 1,500 rounds and 640 at 3,000 and 6,000. \METHOD{} abstains in 93, 81, 74, and 67 percent of replications at the four round counts and violates in 2 of 169 and 6 of 210 selections at 3,000 and 6,000 rounds. Its coverage is 0.997, 0.973, 0.981, and 0.931 with the bias bound (0.97, 0.96, 0.97, 0.93 without).

\section{Candidate generation and reproducibility}
\label{app:design}

The training surfaces are linear in standardized log demand, log median wait, own treatment, and exposure, with one shared exposure coefficient. The Lagrangian path uses $\lambda = 0$ and eleven values doubling from 0.05 to 51.2 (twenty-three log-spaced values on the contagion testbed), the guardrail path $m \in \{1, 1.25, 1.5, 2, 3\}$, and a partial allocation stops after $\max(1, \mathrm{round}(fB))$ units. $B_{\mathrm{boot}} = 10{,}000$, and replication seeds are hashes of setting and index. Every rule above was fixed before the runs.

\end{document}